\documentclass[aps,prb,twocolumn,superscriptaddress,final]{revtex4-2}

\pdfoutput=1

\usepackage[colorlinks=true,allcolors=blue]{hyperref}
\usepackage{graphicx}
\usepackage{amsmath,amssymb,mathrsfs,array,longtable,delarray}
\usepackage{dcolumn}
\usepackage{bm}
\usepackage{booktabs,multirow}

\begin{document}

\title{Effects of biased and unbiased illuminations on dopant-free GaAs/AlGaAs 2DEGs}

\author{A. Shetty}
\affiliation{Institute for Quantum Computing, University of Waterloo, Waterloo N2L 3G1, Canada}
\affiliation{Department of Chemistry, University of Waterloo, Waterloo N2L 3G1, Canada}
\author{F. Sfigakis}
\altaffiliation{corresponding author: francois.sfigakis@uwaterloo.ca}
\affiliation{Institute for Quantum Computing, University of Waterloo, Waterloo N2L 3G1, Canada}
\affiliation{Department of Chemistry, University of Waterloo, Waterloo N2L 3G1, Canada}
\affiliation{Cavendish Laboratory, University of Cambridge, Cambridge CB3 0HE, UK}
\affiliation{Northern Quantum Lights Inc., Waterloo N2B 1N5, Canada}
\author{W. Y. Mak}
\affiliation{Cavendish Laboratory, University of Cambridge, Cambridge CB3 0HE, UK}
\author{K. Das Gupta}
\affiliation{Department of Physics, Indian Institute of Technology Bombay, Mumbai
40007, India}
\author{\\ B. Buonacorsi}
\affiliation{Institute for Quantum Computing, University of Waterloo, Waterloo N2L 3G1, Canada}
\affiliation{Department of Physics and Astronomy, University of Waterloo, Waterloo N2L 3G1, Canada}
\author{M. C. Tam}
\author{H. S. Kim}
\affiliation{Department of Electrical and Computer Engineering, University of Waterloo, Waterloo N2L 3G1, Canada}
\author{I. Farrer}
\affiliation{Cavendish Laboratory, University of Cambridge, Cambridge CB3 0HE, UK}
\affiliation{Department of Electronic and Electrical Engineering, University of Sheffield, Sheffield S1 3JD, UK}
\author{A. F. Croxall}
\author{H. E. Beere}
\affiliation{Cavendish Laboratory, University of Cambridge, Cambridge CB3 0HE, UK}
\author{\\A. R. Hamilton}
\affiliation{School of Physics, University of New South Wales, Sydney NSW 2052, Australia}
\author{M. Pepper}
\affiliation{Department of Electronic and Electrical Engineering, University College London, London WC1E 7JE, UK}
\author{D. G. Austing}
\author{S. A. Studenikin}
\author{A. Sachrajda}
\affiliation{Security and Disruptive Technologies Research Centre, National Research Council of Canada, Ottawa, K1A 0R6, Canada}
\author{\\M. E. Reimer}
\affiliation{Institute for Quantum Computing, University of Waterloo, Waterloo N2L 3G1, Canada}
\affiliation{Northern Quantum Lights Inc., Waterloo N2B 1N5, Canada}
\affiliation{Department of Physics and Astronomy, University of Waterloo, Waterloo N2L 3G1, Canada}
\affiliation{Department of Electrical and Computer Engineering, University of Waterloo, Waterloo N2L 3G1, Canada}
\author{Z. R. Wasilewski}
\affiliation{Institute for Quantum Computing, University of Waterloo, Waterloo N2L 3G1, Canada}
\affiliation{Northern Quantum Lights Inc., Waterloo N2B 1N5, Canada}
\affiliation{Department of Physics and Astronomy, University of Waterloo, Waterloo N2L 3G1, Canada}
\affiliation{Department of Electrical and Computer Engineering, University of Waterloo, Waterloo N2L 3G1, Canada}
\affiliation{Waterloo Institute for Nanotechnology, University of Waterloo, Waterloo N2L 3G1, Canada}
\author{D. A. Ritchie}
\affiliation{Cavendish Laboratory, University of Cambridge, Cambridge CB3 0HE, UK}
\author{J. Baugh}
\altaffiliation{baugh@uwaterloo.ca}
\affiliation{Institute for Quantum Computing, University of Waterloo, Waterloo N2L 3G1, Canada}
\affiliation{Department of Chemistry, University of Waterloo, Waterloo N2L 3G1, Canada}
\affiliation{Northern Quantum Lights Inc., Waterloo N2B 1N5, Canada}
\affiliation{Department of Physics and Astronomy, University of Waterloo, Waterloo N2L 3G1, Canada}
\affiliation{Waterloo Institute for Nanotechnology, University of Waterloo, Waterloo N2L 3G1, Canada}

\begin{abstract}

Illumination is performed at low temperature on dopant-free two-dimensional electron gases (2DEGs) of varying depths, under unbiased (gates grounded) and biased (gates at a positive or negative voltage) conditions. Unbiased illuminations in 2DEGs located more than 70 nm away from the surface result in a gain in mobility at a given electron density, primarily driven by the reduction of background impurities. In 2DEGs closer to the surface, unbiased illuminations result in a mobility loss, driven by an increase in surface charge density. Biased illuminations performed with positive applied gate voltages result in a mobility gain, whereas those performed with negative applied voltages result in a mobility loss. The magnitude of the mobility gain (loss) weakens with 2DEG depth, and is likely driven by a reduction (increase) in surface charge density. Remarkably, this mobility gain/loss is fully reversible by performing another biased illumination with the appropriate gate voltage, provided both \textit{n}\nobreakdash-type and \textit{p}\nobreakdash-type ohmic contacts are present. Experimental results are modeled with Boltzmann transport theory, and possible mechanisms are discussed.
\end{abstract}

\maketitle

\section{Introduction}

Illumination is a well-known technique for increasing the mobility of two-dimensional electron gases (2DEGs) in modulation-doped GaAs/AlGaAs heterostructures at cryogenic temperatures. This effect, known as persistent photoconductivity \cite{NelsonRJ77, Stormer79, Stormer81}, can last for many weeks, as long as the sample is not warmed up above temperature T\,$\sim$\,100 K. In many cases, mobility is increased primarily through the increase of the electron density, which causes more effective Thomas-Fermi screening of charged impurities. The increase in carrier density is mostly achieved by exciting electrons out of deep-level donor impurity complexes known as DX centers \cite{Lang77, Lang79, Kastalsky84,DXcenters1991}. The photo-excited electrons are captured in the GaAs conducting channel. Incremental illumination in small doses (intensity $\times$ duration) can be used for precise tuning of the carrier concentration in modulation-doped ungated heterostructures \cite{YuV18}. Aside from acting on DX centers, illumination may also have other effects, such as activating or deactivating unintentional impurity atoms in the transport channel itself. In modulation-doped 2DEGs, this can be difficult to separate from effects associated with the intentional dopants, which typically outnumber background impurities by three to five orders of magnitude.

The limitation described above can be circumvented by using GaAs-based dopant-free field effect transistors (FET), either in the semiconductor-insulator-semiconductor field effect transistor (SISFET) geometry \cite{Kane93, Saku98, Hirayama98, Kawaharazuka01, Hirayama01, Lilly03, Valeille08} or the heterostructure-insulator-gate field effect transistor (HIGFET) geometry \cite{Harrell99, Willett06, Wendy10, Pan11, ChenJCH12, Croxall13, WangDQ13, Sebastian16, Croxall19}. Dopant-free field effect transistors have been used to produce quantum wires \cite{Harrell99, Reilly01, Klochan06, Sarkozy09-A} and quantum dots \cite{See10, Klochan11, Wendy13, Bogan17, Bogan18}. Relative to their modulation-doped counterparts, dopant-free devices have exceptional reproducibility and low disorder \cite{Sarkozy09-A, Wendy13, See12}, potentially making them suitable to study fragile fractional quantum Hall states \cite{Pan11,Croxall19}. Illumination has been studied in SISFETs, but with conflicting reports \cite{Saku98, Hirayama01, See15}. The effect had not been studied in HIGFETs \footnote{During the peer review of this paper, Ref.\,\onlinecite{FujitaT21} on illumination in HIGFETs was published.}, until very recently \cite{FujitaT21}.

Understanding the effects of biased illumination is particularly relevant for the recently active field of photon spin devices in quantum optoelectronics. Such devices, with possible applications such as spin-to-photon \cite{FujitaT19} or photon-to-spin \cite{HsiaoTK20} conversions, often need to be periodically warmed up to room temperature to be ``reset.'' This issue is also relevant for some proposed single photon source proposals \cite{Blumenthal07,Brandon21} which would use so-called ``lateral'' dopant-free p-i-n junctions \cite{Dai13,Dai14,ChungYC19}.

In this article, we report on the effects of illumination on 2DEGs in dopant-free Hall bars in the HIGFET geometry, with the 2DEG depth below the surface ranging from 30 nm to 310 nm. We model the mobility, and quantify the effects of illumination on surface states and background impurities. We characterize biased illumination, an experimental technique where illumination is performed while gates are held at finite voltages, whose effects are markedly different from unbiased illumination. Biased illumination appears to allow \textit{in-situ} control of surface charge. Section II covers the growth and fabrication of samples, section III covers the scattering theory used to model mobilities, section IV covers the transport experiments, and section V covers the discussion and conclusions about the work presented here.

\begin{table}[h]
    \caption{Index of samples for which data is shown in at least one figure of the main text (20 devices in total were measured). The MBE layer structure of the wafers is shown in Figure \ref{Fig1}.}
    \begin{ruledtabular}
    \begin{tabular}{ccccc}
    ~~~~~~ & Sample & Wafer & 2DEG & AlGaAs \\
    ~~~~~~ & ID & ID & depth (nm) & barrier (nm) \\
    \hline ~\\
    Series I & A & W639 & 160 & 150 \\
    ~~~~~~~~ & B & W640 & 110 & 100 \\
    ~~~~~~~~ & C & W641 & ~\,60 & ~\,50 \\
    ~~~~~~~~ & D & V627 & ~\,30 & ~\,20 \\ ~\\
    Series II & E & G404 & 310 & 300 \\ 
    ~~~~~~~~  & F & G404 & 310 & 300 \\ 
    ~~~~~~~~  & G & G404 & 310 & 300 \\ 
    ~~~~~~~~  & H & G373 & 160 & 150 \\ 
    ~~~~~~~~  & J & G372 & 110 & 100 \\ 
    ~~~~~~~~  & K & G370 & ~\,75 & ~\,65 \\ 
    ~~~~~~~~  & L & G370 & ~\,75 & ~\,65 \\ 
    ~~~~~~~~  & M & G370 & ~\,75 & ~\,65 \\ 
    \end{tabular}
    \end{ruledtabular}
    \label{tab:index}
\end{table}

\begin{figure}[t]
    \includegraphics[width=0.75\columnwidth]{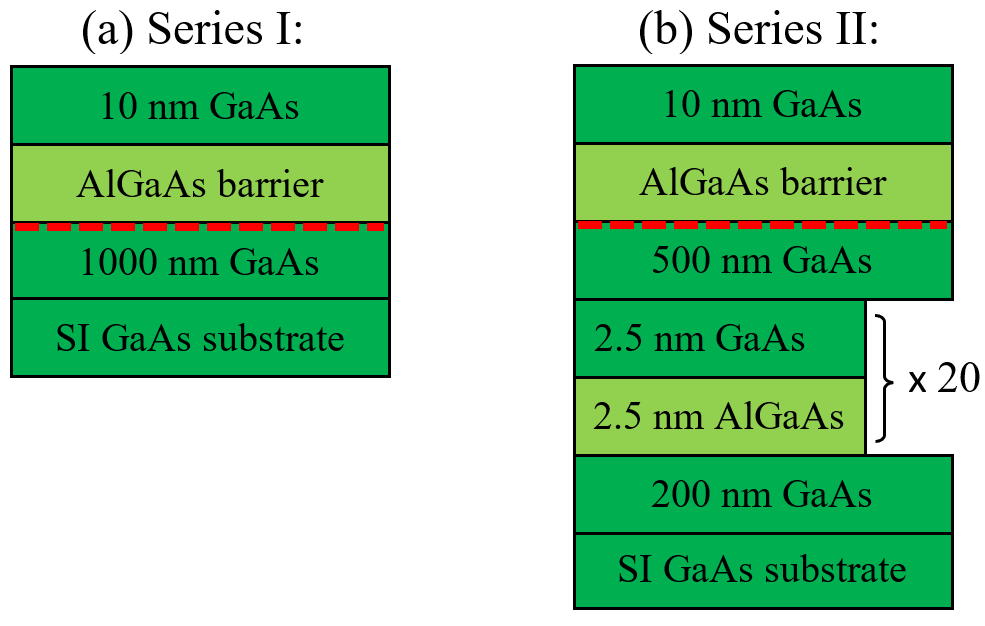}
    \caption{(color online) MBE layer structure of GaAs/AlGaAs single heterojunctions used in: (a) Series I, and (b) Series II. Table \ref{tab:index} lists the AlGaAs barrier thickness for each wafer. The dashed red line indicates the location of the 2DEG.}
    \label{Fig1}
\end{figure}  

\begin{figure}[b]
    \includegraphics[width=0.80\columnwidth]{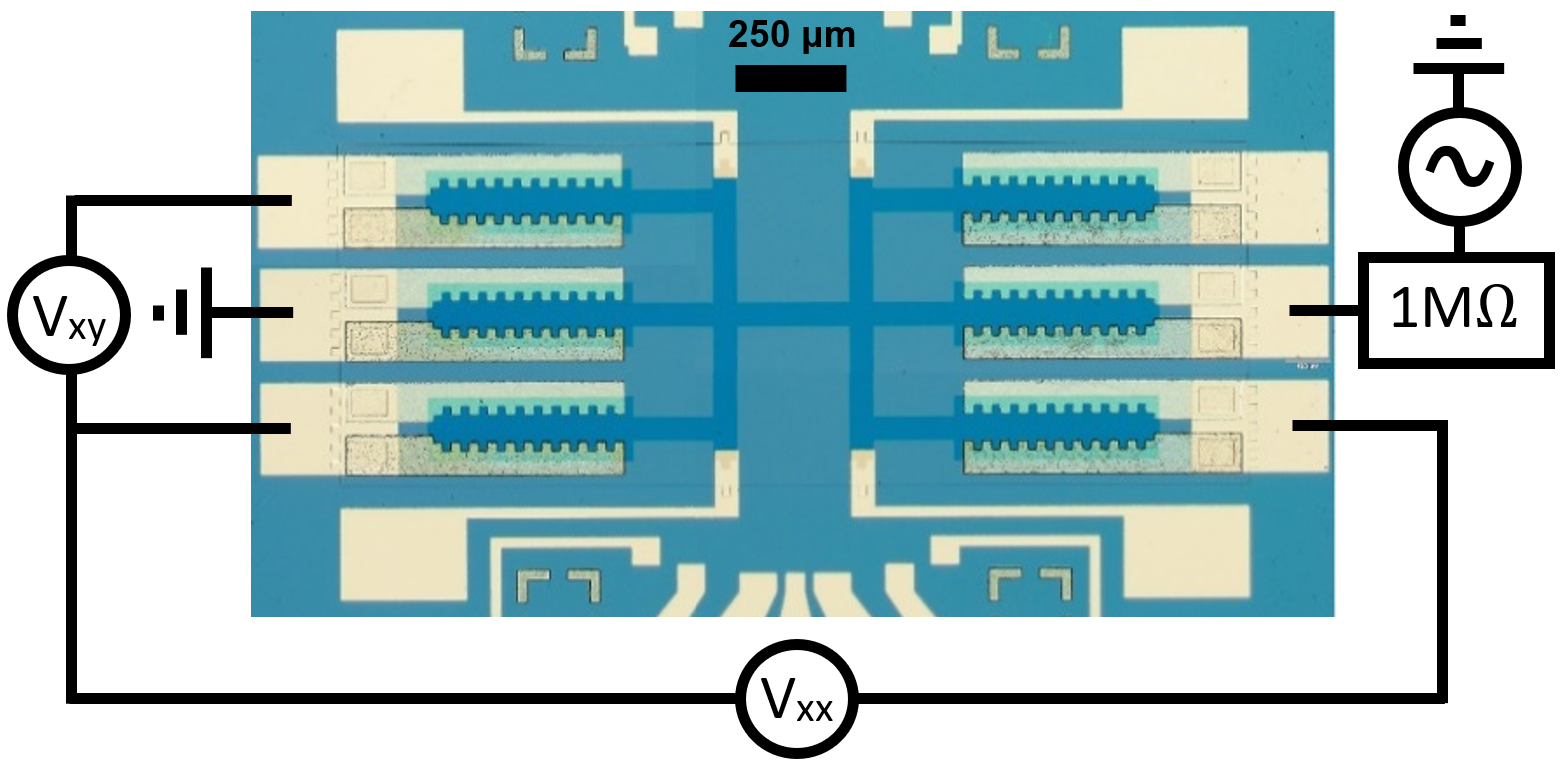}
    \caption{(color online) (a) Optical photograph of an ambipolar Hall bar with a semi-transparent 5 nm Ti topgate (in dark blue), with schematic of the circuit used for measurements.}
    \label{Fig2}
\end{figure}  

\section{Sample growth and fabrication}

Two series of dopant-free GaAs/AlGaAs heterostructures \footnote{Wafers from Series I (Series II) were grown at the University of Cambridge (University of Waterloo).}, each with varying
single-heterojunction depths, were grown by molecular beam epitaxy (MBE). Series I included
three GaAs/Al$_{0.33}$Ga$_{0.67}$As structures (W639, W640, and W641) grown on the same day, and an ultra-shallow heterojunction (V627) grown in a different chamber with an Al$_{0.90}$Ga$_{0.10}$As barrier. Series II, grown later in a third chamber, included three GaAs/Al$_{0.30}$Ga$_{0.70}$As heterojunctions (G370, G372, and G373) grown over two consecutive days, as well as a much deeper heterojunction (G404) grown several weeks later. The MBE layer structures of the two series are shown in Figure 1. The AlGaAs barrier layer thickness was varied from 20 nm to 300 nm, and is listed in Table \ref{tab:index} for each wafer. All wafers were grown on 3" semi-insulating (SI) GaAs (001) substrates.

Hall bars were fabricated on all 8 wafers, and oriented in the high mobility crystal direction $[1\bar{1}0]$. All devices from Series I wafers were unipolar Hall bars (2DEG only). All devices from Series II wafers (except where specified) were ambipolar Hall bars, with both
\textit{n}\nobreakdash-type and \textit{p}\nobreakdash-type ohmic contacts. This allowed a two-dimensional electron gas (2DEG) or a two-dimensional hole gas (2DHG) to be induced, depending on the voltage polarity applied to the topgate.

The fabrication of unipolar (2DEG only) Hall bars on wafers from Series I is extensively described in Ref.\,\cite{Wendy10}. Briefly, after deposition and anneal of the recessed Ni/AuGe/Ni \textit{n}\nobreakdash-type ohmic contacts, a 500 nm insulator layer of photoimageable polyimide (HD4104) was spin-coated and cured at 250$^\circ$C, with typical breakdown voltages of 25$-$35 Volts [see Fig.\,1(c) in Ref.\,\onlinecite{Wendy13}]. Above the insulator layer, a thin Ti/Au (5 nm/1 nm) semi-transparent topgate covers the entire surface of the 2DEG (overlapping the ohmic contacts), and varies the electron density. Surprisingly, otherwise identical Hall bars with a thicker, ``opaque'' topgate (Ti/Au 20/80 nm) gave similar results as those presented here with the thin topgates. Because of the thick polyimide insulator layer, we speculate that light can travel inwards and underneath the topgate from its edges.

The fabrication of ambipolar Hall bars on wafers from Series II is described in Refs.\,\cite{Wendy10} and \cite{ChenJCH12}; such a device is shown in Fig.\,\ref{Fig2}. Briefly, after the deposition/anneal of recessed Ni/AuGe/Ni \textit{n}\nobreakdash-type ohmic contacts and recessed AuBe \textit{p}\nobreakdash-type ohmic contacts, a 300 nm SiO$_2$ layer was deposited by plasma-enhanced chemical vapor deposition (PECVD). Above the insulator layer, a Ti/Au topgate covers the entire surface of the 2DEG (overlapping the ohmic contacts), and varies the carrier density. As with Series I, devices with a 5 nm semi-transparent Ti topgate gave similar results to those with a Ti/Au topgate.

Overall, 20 Hall bars (with two types of gate dielectrics, SiO$_2$ and polyimide) were fabricated from 8 GaAs/AlGaAs wafers grown in three MBE growth chambers located in Canada and the UK. The Appendix shows supporting data from additional Hall bars in another 6 wafers, from two more MBE growth chambers.

\section{Boltzmann Transport Model}
\label{section:theory}

The mobility $\mu$ of carriers (electrons or holes) is limited by their interactions with their environment via scattering events, and relates to the momentum relaxation time $\tau$ (also known as the transport scattering time) by $\mu = e\tau/m^*$, where $m^*$ is the effective mass and $e$ the elementary charge. The total scattering rate $1/\tau_{total}$ of carriers is simply the sum of the rates of all scattering mechanisms occurring in the system,
$\frac{1}{\tau_{total}}=\sum_{i} \frac{1}{\tau_{i}}$ (Matthiessen's rule \cite{Matthiessen1864}). Thus, we model the mobility as \cite{Ihn2010book}:
\begin{equation}
\frac{1}{\mu}=\frac{m^*}{e}
\left(\frac{1}{\tau_{\textsc{bi-1}}}
+\frac{1}{\tau_{\textsc{bi-2}}}
+\frac{1}{\tau_{\textsc{ir}}}+\frac{1}{\tau_{\textsc{sc}}} \right) \label{eq:Matthiesen}
\end{equation}
\noindent where $1/\tau_{\textsc{bi-1}}$ is the scattering rate due to ionized background impurities in AlGaAs,  $1/\tau_{\textsc{bi-2}}$ is the scattering rate due to ionized background impurities in GaAs, $1/\tau_{\textsc{ir}}$ is the scattering rate due to the GaAs/AlGaAs interface roughness, and $1/\tau_{\textsc{sc}}$ is the scattering rate due to surface charges. Phonon scattering is neglected, as all measurements were performed at the same low temperature (T $\sim$ 1.5 K) and are only compared relative to each other. Sources of scattering are treated within the semi-classical Boltzmann transport formalism. A detailed derivation of how each scattering mechanism contributes to the mobility can be found elsewhere \cite{Ando82-A, Davies1998book, Gold88}, but some of the key approximations and expressions are described below and in the Appendix.

\begin{figure}[t]
    \includegraphics[width=0.40\columnwidth]{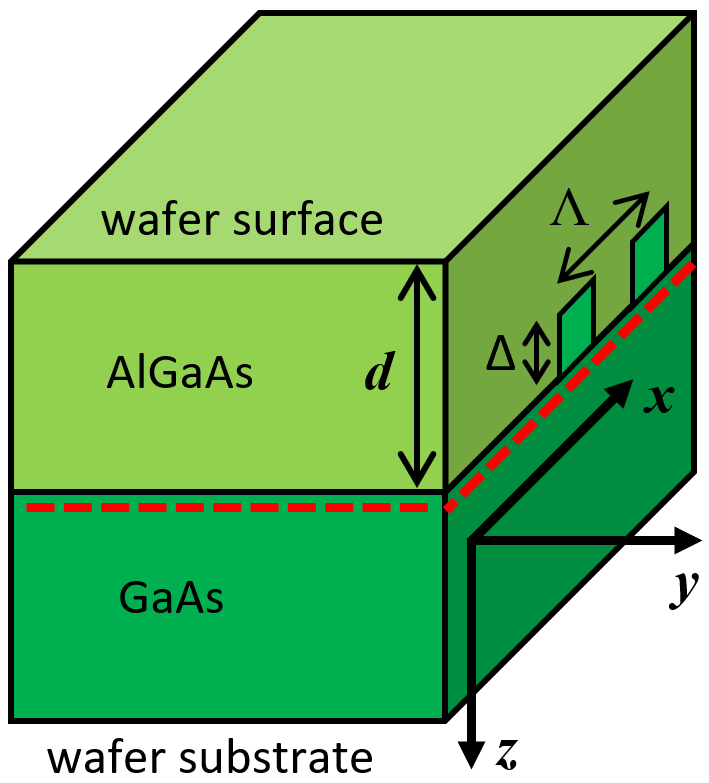}
    \caption{(color online) Heterostructure layout for theory model. The dashed red line indicates the location of the 2DEG (at $z=0$), at a depth $d$ below the wafer surface. Interface roughness irregularities are shown, along with definitions for their separation distance $\Lambda$ and average height $\Delta$. The GaAs layer is treated as semi-infinite (towards the substrate), and the 10 nm GaAs cap layer [Fig.\,\ref{Fig1}] is treated as part of the AlGaAs barrier.}
    \label{Fig3}
\end{figure}

Electrons are described by the Fang-Howard wavefunction $\Psi{(\textbf{r},z)} \propto \psi(z)~e^{i\textbf{k$\cdot$r}}$, where \textbf{r} is any direction within the $x$-$y$ (2DEG) plane and $z$ is the MBE growth direction. The 2DEG resides at the GaAs/AlGaAs interface at $z$\,=\,0 [Fig.\,\ref{Fig3}]. Following the orientation convention in Fig.\,\ref{Fig3}, the wavefunction $\psi(z)$ is \cite{Fang66,SternHoward67}:
\begin{eqnarray}
\psi(z)=0  \qquad\qquad\qquad\qquad \text{for}~z<0 \\
\psi(z)=\left(\frac{b^{3}z^2}{2}\right)^{1/2}e^{-bz/2}~\,~ \text{for}~z \geqslant 0\\
\text{with}\quad b=\left(\frac{33m_{z}e^{2}n_{\textsc{2d}}}{8\hbar^2\epsilon_0\epsilon_r}\right)^{1/3}
\end{eqnarray}
where $m_z$ is the effective mass in the growth direction ($m_z =  m^* = 0.067 m_0$ for electrons with $m_0$ the free electron mass), $\epsilon_0$ the vacuum permittivity, $\epsilon_r$ the relative permittivity of GaAs and AlGaAs (approximating $\epsilon_r=\epsilon^{\text{GaAs}}_{r} \approx \epsilon^{\text{AlGaAs}}_r \approx 12.8 $), and $n_{\textsc{2d}}$ the 2D carrier sheet density. The $\psi(z)$ wavefunction typically spans 10-30 nm at the carrier densities used in experiments, and its maximum occurs at a distance of $2/b$ below the GaAs/AlGaAs interface. The Fang-Howard wavefunction leads to the following form factor $F_\textsc{fh}(q)$:
\begin{eqnarray}
F_\textsc{fh}(q)&=&\int_0^\infty \int_0^\infty |\psi(z)|^2 ~|\psi(z')|^2~e^{-q|z-z'|}~dz~dz'\quad\quad\\
F_\textsc{fh}(q)&=&\frac{2}{8}\left(\frac{b}{b+q}\right)^3 + \frac{3}{8}\left(\frac{b}{b+q}\right)^2 + \frac{3}{8}\left(\frac{b}{b+q}\right)
\end{eqnarray}
\noindent where $q=2k_F \sin(\theta/2)$ is the scattering wavevector (with scattering angle $\theta$) and $k_F=\sqrt{2\pi n_{\textsc{2d}}}$ is the Fermi wavevector.

Taking into account that the potential from an ionized impurity is partially screened by the 2DEG (dielectric screening) and using the Thomas-Fermi approximation, the dielectric function $\epsilon(q)$ can be written as:
\begin{equation}
\epsilon(q)=1+\frac{e^2 }{ 2\epsilon_0\epsilon_rq}\frac{m_z}{\pi\hbar^2}
F_\textsc{fh}(q)
\end{equation}
which includes $F_\textsc{fh}(q)$ to account for the finite width of the 2DEG wavefunction.

Applying Fermi's golden rule to a 2DEG with scattering potential $U(q)$, the following general expression for the resulting scattering rate at temperature $T=0$ is obtained \cite{Gold88, MacLeod09}:
\begin{equation}
\frac{1}{\tau}=\frac{m^*}{\pi \hbar^3 k^2_F}~\int^{2k_F}_0 \frac{|U(q)|^2}{\epsilon(q)^2}\frac{q^2}{\sqrt{4k^2_F-q^2}}~dq
\label{eq:FermiGoldenRule}
\end{equation}
for which the corresponding $|U(q)|^2$ terms and associated scattering rates $1/\tau_{\textsc{ir}}$, $1/\tau_{\textsc{sc}}$, $1/\tau_{\textsc{bi-1}}$, and $1/\tau_{\textsc{bi-2}}$ are described respectively by equations (\ref{eq:1byTauIR}), (\ref{eq:1byTauSC}), (\ref{eq:1byTauBI-AlGaAs}), and (\ref{eq:1byTauBI-GaAs}) in the Appendix. For convenience, only the final expressions for the scattering rate of each mechanism are listed below:
\begin{eqnarray}
\frac{1}{\tau_{\textsc{ir}}}&=&
\frac{(\Lambda\Delta)^2\,m^*}{2\hbar^3 k_F^2}
\left(\frac{n_{\textsc{2d}}e^2}{2\epsilon_0\epsilon_r}\right)^2
\int_0^\pi \frac{q^2\,e^{-q^2\Lambda^2/4}}{\epsilon(q)^2}~d\theta
\label{eq:1byTauIRfinal}\\
\frac{1}{\tau_{\textsc{sc}}}&=&\frac{N_{\textsc{sc}}\,m^*}{2\pi\hbar^3 k_F^2}
\left(\frac{e^2}{2\epsilon_0\epsilon_r}\right)^2
\int_0^\pi \frac{e^{-2q|d|}}{\epsilon(q)^2\,(1+q/b)^6}~d\theta \qquad \label{eq:1byTauSCfinal}\\
\frac{1}{\tau_{\textsc{bi-1}}}&=&
\frac{N_{\textsc{bi-1}}\,m^*}{2\pi \hbar^3 k^2_F}
\left(\frac{e^2}{2\epsilon_0\epsilon_r}\right)^2
\int^\pi_0\frac{(1+q/b)^{-6}}{2q\,\epsilon(q)^2}\,d\theta
\label{eq:1byTauBI-AlGaAsfinal}\\
\frac{1}{\tau_{\textsc{bi-2}}}&=&\frac{N_{\textsc{bi-2}}\,m^*}{2\pi \hbar^3 k^2_F}
\left(\frac{e^2}{2\epsilon_0\epsilon_r}\right)^2
\int^\pi_0 \frac{F_{\text{\tiny GaAs}}(q)}{\epsilon(q)^2}\,d\theta~. \label{eq:1byTauBI-GaAsfinal}
\end{eqnarray}
\noindent where $\Delta$ is the height of irregularities in the $z$ direction at the GaAs/AlGaAs interface (Fig.\,\ref{Fig3}), $\Lambda$ is the separation distance in the $x$-$y$ plane between these irregularities [Fig.\,\ref{Fig3}], $d$ is the distance of the GaAs/AlGaAs interface (or nominal 2DEG depth) to the wafer surface, $N_{\textsc{sc}}$ is the sheet concentration of surface charges, $N_{\textsc{bi-1}}$ is the volume concentration of background impurities in AlGaAs,  $N_{\textsc{bi-2}}$ is the volume concentration of background impurities in GaAs, and $F_{\text{\tiny GaAs}}(q)$ is a form factor described by equation (\ref{eq:FormFactor-GaAs}). By substituting the scattering rates $1/\tau_i$ expressed in eqns.\,(\ref{eq:1byTauIRfinal})$-$(\ref{eq:1byTauBI-GaAsfinal}) into equation (\ref{eq:Matthiesen}), the transport mobility can be calculated.

This model has been very successful at describing behavior both from shallow and deep dopant-free 2DEGs [see Ref.\,\onlinecite{Wendy10}, as well as Figures \ref{FigAppndx-SCmobility} and \ref{FigAppndx-MBEcleanup} in the Appendix]. The model's  implementation in code is available online \footnote{The computer program written in C and a fitting routine written for MATLAB\textregistered ~is available at https://github.com/arjuntalk/2DEG-mobility-base-model}. Since we experimentally extract the interface roughness parameters $\Delta$ and $\Lambda$ from wafer surface analysis with an atomic force microscope (AFM), these parameters are not unrestricted free variables when fitting experimental mobilities to this model. In the case of deep 2DEGs (where $N_{\textsc{sc}}$ is negligible), curve-fitting can be reduced to a single free variable: the average background impurity concentration $\overline{N_{\textsc{bi}}} = N_{\textsc{bi-1}} = N_{\textsc{bi-2}}$ [see Figure \ref{FigAppndx-MBEcleanup} in the Appendix].

\section{Experiments and analysis}

The following applies to all Hall bar measurements described here. Constant current (100 nA) four-terminal measurements were performed in two pumped-{$^4$}He cryostats (T\,$\sim$\,1.5 K), with standard AC lock-in techniques using SR-830 lock-ins and SR-560 voltage pre-amplifiers \footnote{Hall bars from Series I (Series II) were measured at the University of Cambridge (University of Waterloo)}. Typical ohmic contact resistances were 500-1500 $\Omega$ in 2DEGs (these are resistive because of the thick 120 nm Ni capping layer) and less than 200 $\Omega$ in 2DHGs. There was no measurable leakage current from the topgate to the 2DEG above the $\sim$\,10 pA noise floor of the DC measurement setup, for any topgate voltage applied. Mobility and carrier density were obtained from the following relations:
\begin{eqnarray}
n_{\textsc{2d}} = \frac{IB}{eV_{\textsc{h}}} \quad \\
\mu = \frac{I(L/W)}{e n_{\textsc{2d}} V_{xx}}
\label{eq:exp-densitymobility}
\end{eqnarray}
\noindent where $n_{\textsc{2d}}$ is the Hall electron carrier density, $I$ is the ac excitation current (along the $x$ direction), $B$ is the magnetic field (oriented perpendicular to the 2DEG plane), $V_{\textsc{h}}$ is the Hall voltage (obtained from $V_{\textsc{h}} = [V_{xy}(B)-V_{xy}(-B)]/2$, which eliminates any offsets in $V_{xy}$ at $B$=0), $W$ is the width of the Hall bar (corresponding to the edges of the topgate), $L$ is the (center-to-center) distance between voltage probe contacts on the Hall bar, and $V_{xx}$ is the voltage drop along the direction of the ac current $I$ in the high-mobility crystal direction $[1\bar{1}0]$. Data for carrier density and mobilities was taken with four significant digits and uncertainty ranging from $\pm$0.05\% to $\pm$0.3\%. Error bars in plots are thus smaller than the marker symbols used, and are not shown. All data shown in Figures \ref{Fig4}$-$\ref{Fig6} and  \ref{FigNEQ0}$-$\ref{FigUni} have been reproduced in at least two Hall bars, unless noted otherwise.

Experimental results are presented in two parts: unbiased illumination in section \ref{section:Vtg-zero}, and biased illumination in section \ref{section:Vtg-neq0}.

\subsection{Unbiased illumination, while $V_{\text{topgate}}$\,=\,0}
\label{section:Vtg-zero}

\begin{figure}[t]
    \includegraphics[width=\columnwidth]{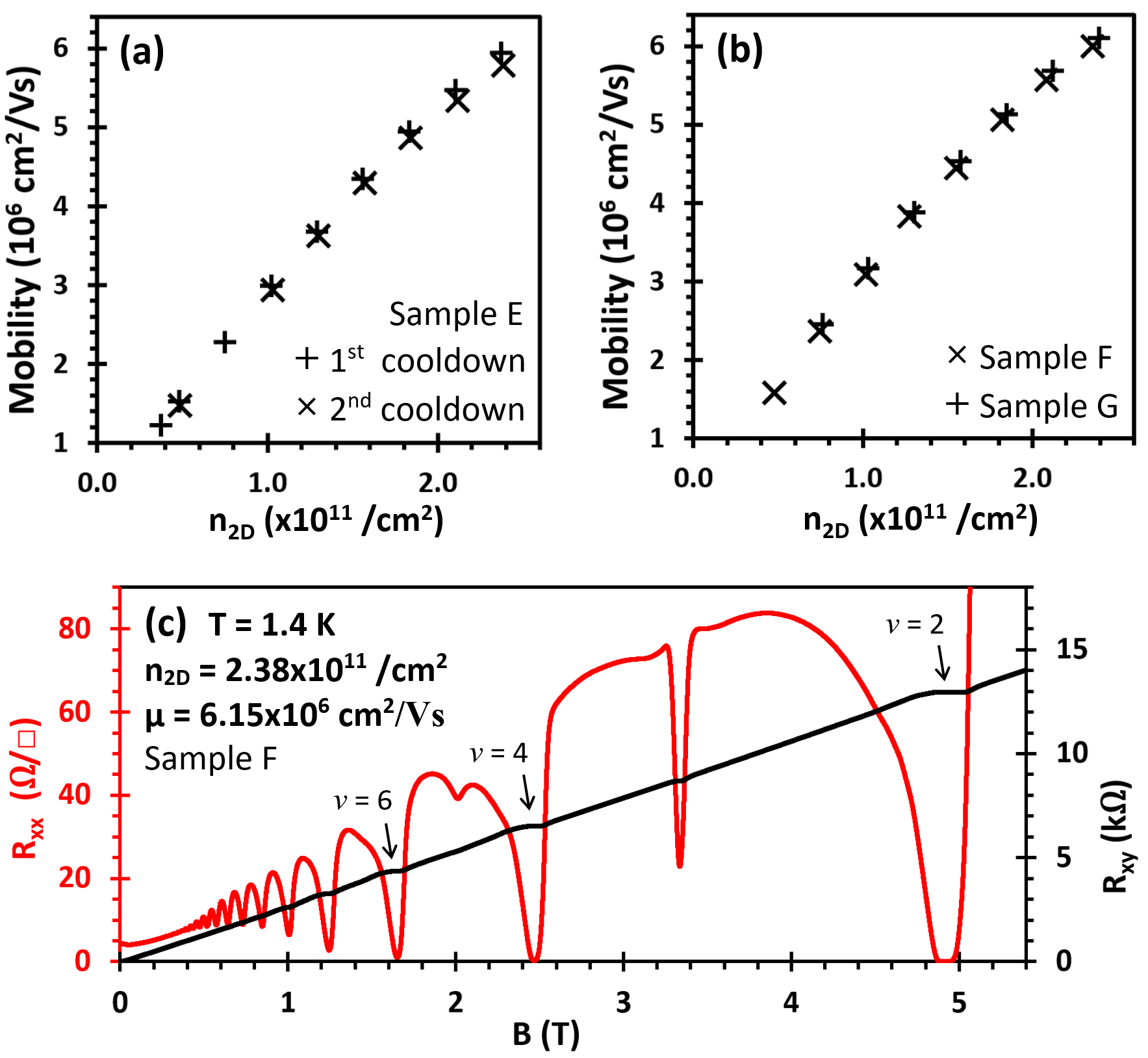}
    \caption{(color online) (a) Reproducibility of mobility between two cooldowns of the same Hall bar, before illumination. (b) Reproducibility of mobility characteristics for two Hall bars from the same wafer before illumination. (c) Typical quantum Hall effect and Shubnikov-de-Haas oscillations before illumination, with visible quantized Hall plateaus at filling factors $\nu = 2, 3, 4, 6,$ and 8.}
    \label{Fig4}
\end{figure}

Examples of reproducibility prior to illumination are shown in Figure \ref{Fig4}(a), showing mobility measurements from two separate cooldowns on the same Hall bar, and in Figure \ref{Fig4}(b), showing  mobility measurements on two separate Hall bars from the same wafer. The narrow Shubnikov-de-Haas (SdH) oscillations and quantum Hall (QH) effect observed in Figure \ref{Fig4}(c) are consistent with high mobilities. The minima of SdH oscillations reach $R_{xx}$\,=\,0; there is no parallel conduction. The carrier density extracted from the SdH oscillations matches that of the Hall density; the 2DEG occupies a single subband.

\begin{figure}[t]
    \includegraphics[width=\columnwidth]{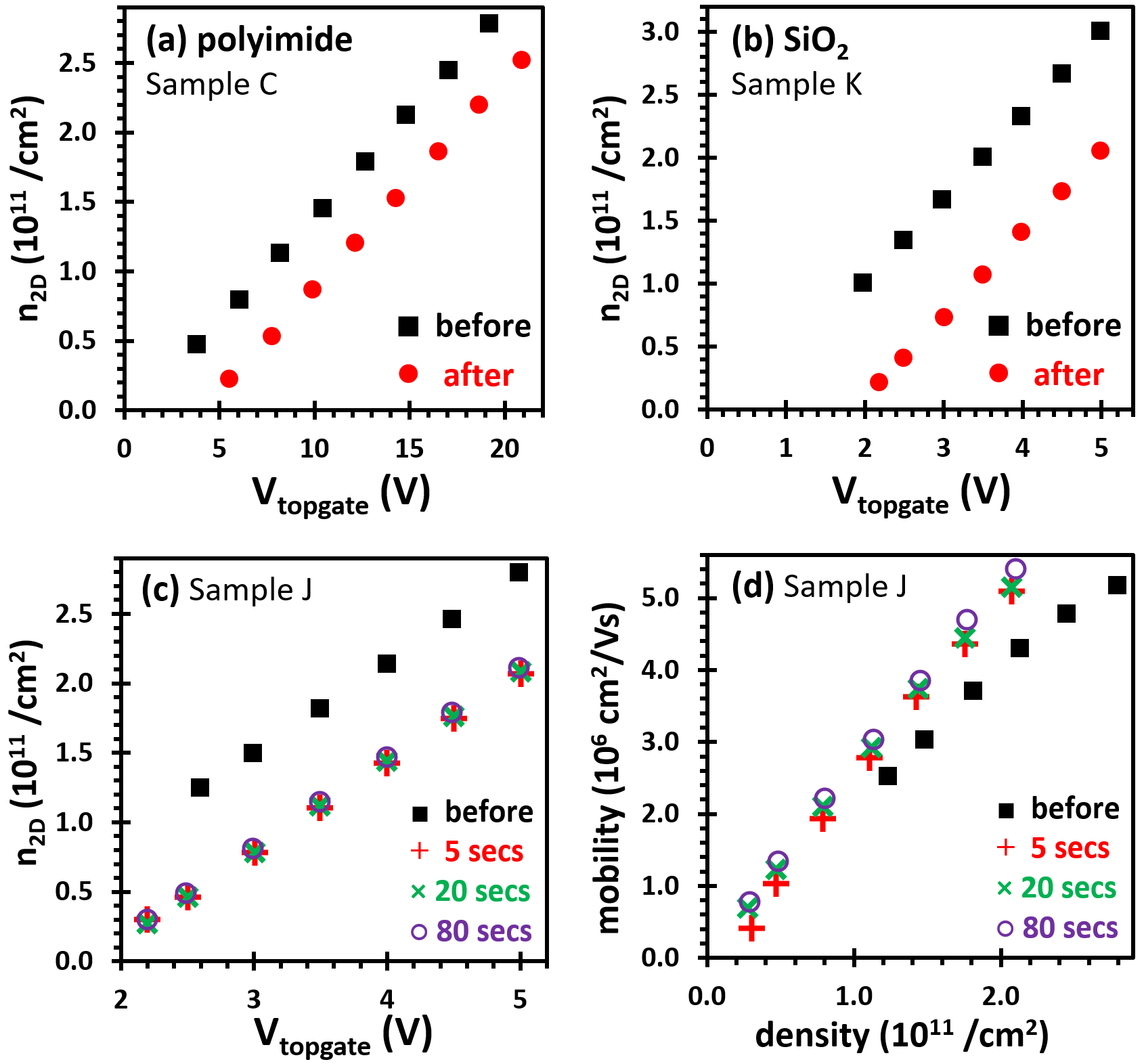}
    \caption{(color online) Electron density versus topgate voltage relationships $n_{\textsc{2d}}(V_{\text{topgate}})$, before (black squares) and after (red circles) illumination, typical of Hall bars from all wafers fabricated with a: (a) polyimide insulator, and (b) SiO$_2$ insulator. (c) In all wafers, after the initial 5 seconds, longer illuminations did not appear to cause further change in $n_{\textsc{2d}}(V_{\text{topgate}})$. (d) After illumination, electron mobility improves. Most of the change occurs during the first 5 seconds of illumination, with eventual saturation at longer illumination times.}
    \label{Fig5}
\end{figure}

For all Hall bars from both Series I (polyimide insulator) and Series II (SiO$_2$ insulator), the $n_{\textsc{2d}}(V_{\text{topgate}})$ relationships shown in Figures \ref{Fig5}(a) and \ref{Fig5}(b) are linear and non-hysteretic, confirming no significant gate leakage or re-chargeable traps in the insulator. Before illumination, the topgate voltage in samples from Series II was limited to $|V_{\text{topgate}}| \leqslant 5$~V to prevent gate hysteresis (this is discussed further in section \ref{section:Vtg-neq0}). Series I samples were illuminated directly with a red LED driven at 10 mA. Series II samples were illuminated indirectly with another red LED driven at 78 mA. Illumination ranged in durations from 5 seconds to 8 minutes, while samples were grounded and no voltage was applied to the topgate (unbiased illumination). The largest change occurs during the first 5 seconds of illumination for Series II samples, and subsequent illuminations have a smaller effect. A typical example for the density-topgate relation $n_{\textsc{2d}}(V_{\text{topgate}})$ is shown in Figure \ref{Fig5}(c), before and after illuminations of varying durations.

Two observations can be drawn from Figures \ref{Fig5}(a) and \ref{Fig5}(b). First, the slope of the density-voltage relation, a direct measurement of the capacitance between the 2DEG and the topgate, does not change before/after illumination within 1-2\%. Second, it becomes more difficult to induce electrons in the GaAs channel after illumination, irrespective of insulator type or MBE growth chamber, requiring significantly higher 2DEG turn-on threshold topgate voltages $V_{\text{th}}$, defined as the extrapolated $n_{\textsc{2d}}$\,=\,0 intercept on the topgate voltage axis \footnote{This definition minimizes variations in turn-on voltages between individual ohmic contacts, related to fabrication parameters rather than the GaAs/AlGaAs material itself.}. This is contrary to what was observed in previous studies of illumination on 2DEGs in SISFETs \cite{Saku98, Hirayama01, See15}, where $V_{\text{th}}$ was lower to achieve the same electron density after illumination. However, in SISFETs, the gate used to induce a 2DEG is a degenerately-doped GaAs cap layer, which completely screens surface states and prevents them from affecting $V_{\text{th}}$. With metal gates in dopant-free HIGFETs, such screening does not occur. Charged traps inside or at the interface between the amorphous gate dielectric (SiO$_2$ or polyimide) and the GaAs cap layer can affect $V_{\text{th}}$, as well as any surface treatment immediately prior to gate dielectric deposition \cite{FujitaT21}.


A typical example of the mobility-density relation $\mu(n_{\textsc{2d}})$ is shown in Figure \ref{Fig5}(d) at $T=1.4$~K, before and after illuminations of varying durations. After illumination, electron mobility is improved. Most of the change occurs during the first 5 seconds (15 seconds) of illumination for Series II (Series I) samples, with eventual saturation at longer illumination times.

\begin{figure*}[t]
    \includegraphics[width=2.0\columnwidth]{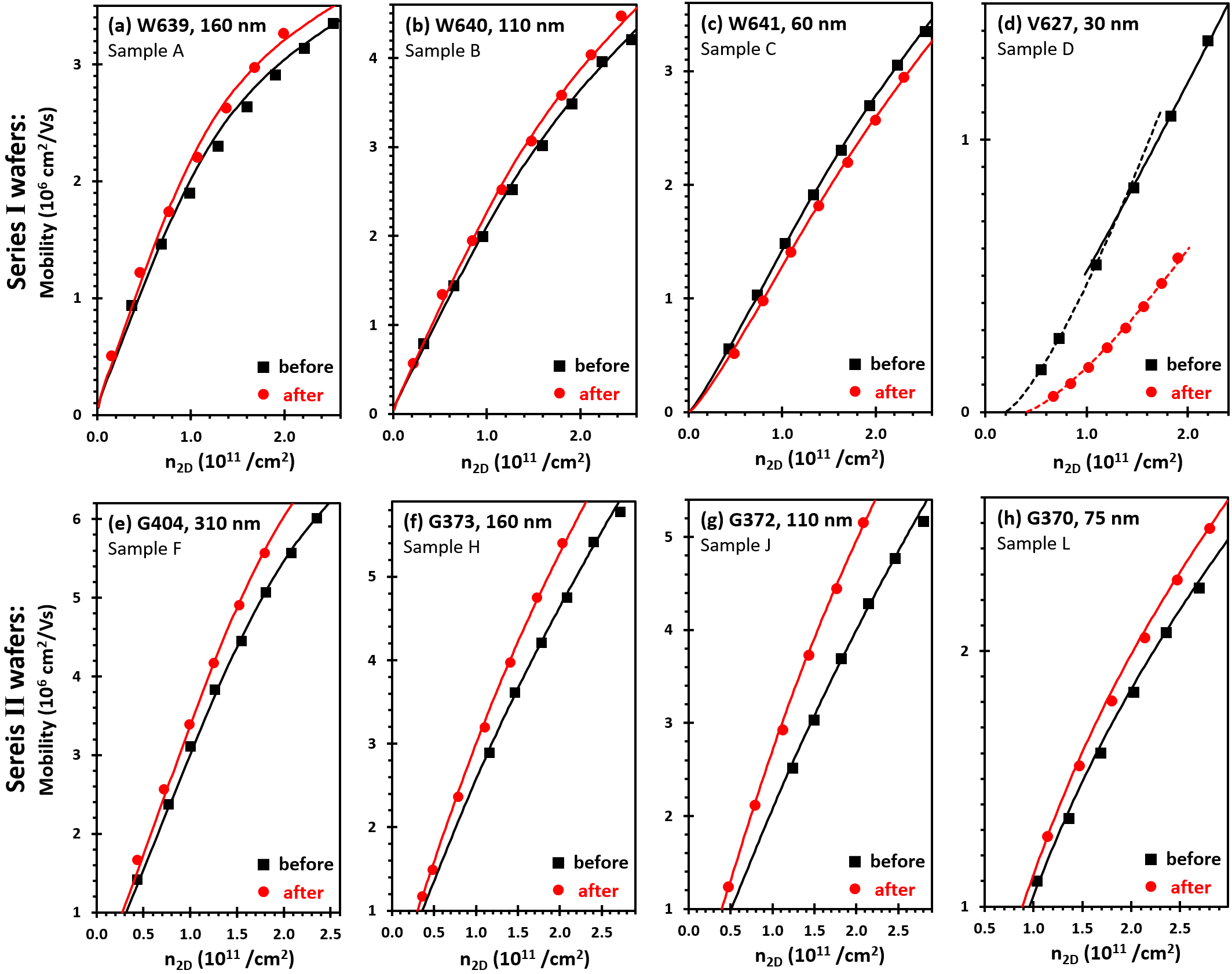}
    \caption{(color online) Electron mobilities before (black squares) and after (red circles) illumination for: (a)$-$(d) Series I wafers and (e)$-$(h) Series II wafers. Series I (Series II) wafers with a polyimide (SiO$_2$) insulator were illuminated for 4 minutes (80 seconds). Solid lines are fits to the Boltzmann transport model described in section \ref{section:theory}, equation (\ref{eq:Matthiesen}) and equations (\ref{eq:1byTauIRfinal})$-$(\ref{eq:1byTauBI-GaAsfinal}) with fit parameter values listed in Table \ref{tab:Fits}. Dashed lines (red or black) in panel (d) are fits to equation (\ref{eq:percolation}), describing transport near the 2D percolation threshold.}
    \label{Fig6}
\end{figure*}  

Figure \ref{Fig6} shows experimental electron mobilities as a function of electron density for wafers from Series I and Series II, before and after illumination to saturation. As expected, within a wafer series, mobility decreases as the 2DEG becomes closer to the surface \footnote{Atomic force microscopy analysis revealed wafer W640 had an unexpectedly larger interface roughness than the other two wafer of the series, accounting for its lower-than-expected mobility.}, in line with previous studies, both in dopant-free 2DEGs and modulation-doped 2DEGs \cite{Wendy10, Wendy13}. Scattering from surface charges is the primary mechanism for this mobility degradation, and becomes pronounced for 2DEG depths smaller than $\sim$\,80 nm \cite{Kawaharazuka00, Hirayama01, Wendy10}. This is also shown experimentally and theoretically in Fig.\,\ref{FigAppndx-SCmobility} in Appendix \ref{Appndx:SurfaceCharge}. In Figure \ref{Fig6}, illumination increases the electron mobility by up to 30\% for the six deepest 2DEGs, where the surface is 75 nm or more away. For the two shallowest 2DEGs (Figs.\,\ref{Fig6}(c) and \ref{Fig6}(d), 60 nm and 30 nm deep, respectively), mobility decreases after illumination. This decrease in only the two shallowest of the eight 2DEGs surveyed strongly suggests that surface charge plays a role. Previous studies on SISFETs used 2DEGs deep below the surface ($d$\,=\,250 nm in Ref.\,\cite{Saku98}, 150 nm in Ref.\,\cite{Hirayama01}, and 185 nm in Ref.\,\cite{See15}). Their observations of increased post-illumination mobility on dedicated Hall bars are consistent with the data shown here. Likewise, a mobility increase after illumination was also observed in the other illumination study on HIGFETs \cite{FujitaT21}, with their 2DEG located 115 nm below the surface.

\begin{table*}[t]
    \caption{List of parameters used to model the 2DEG mobilities shown in Figure \ref{Fig6}, before and after illumination (dark/light): $d$ is the 2DEG depth below the wafer surface, $\Delta$ is the average height of surface irregularities, $\Lambda$ is the average separation between these irregularities, $\overline{N_{\textsc{bi}}}$ is the average concentration of background impurities in GaAs and AlGaAs (with $\overline{N_{\textsc{bi}}} = N_{\textsc{bi-1}} = N_{\textsc{bi-2}}$), $N_{\textsc{sc}}$ is the sheet density of surface charges, and $\Delta V_{\text{th}}$ is the change in the 2DEG threshold voltage before/after illumination, rounded to the first digit after the decimal point. The parameters $d$, $\Delta$, and $\Lambda$ do not change between before/after illumination.}
    \begin{ruledtabular}
    \begin{tabular}{lccccccccccc}
    & & &
    & & \multicolumn{2}{c}{before illumination}
    & & \multicolumn{2}{c}{after illumination} & & \vspace{1 mm} \\
    Wafer & $d$ & $\Delta$ & $\Lambda$
    & & $\overline{N_{\textsc{bi}}}$ & $N_{\textsc{sc}}$
    & & $\overline{N_{\textsc{bi}}}$ & $N_{\textsc{sc}}$ & & $\Delta V_{\text{th}}$  \\
    ~~\,ID & (nm) & (nm) & (nm)
    & & (cm$^{-3})$ & (cm$^{-2})$
    & & (cm$^{-3})$ & (cm$^{-2})$ & & (Volt)\footnotemark[1] \\ \hline
     V627\footnotemark[2] & ~\,30 & 0.11 & 15
    & & 6.8$\times$10$^{13}$ & 2.5$\times$10$^{11}$
    & & \textendash & \textendash & & \textendash \\~\\
    W639 & 160 & 0.15 & 14
    & & 1.3$\times$10$^{14}$ & $<$1$\times$10$^{10}$
    & & 1.2$\times$10$^{14}$ & $<$1$\times$10$^{10}$ & & +0.9 \\
    W640 & 110 & 0.11 & 14
    & & 1.3$\times$10$^{14}$ & 0.2$\times$10$^{11}$
    & & 1.2$\times$10$^{14}$ & 0.3$\times$10$^{11}$ & & +1.7 \\
    W641 & ~\,60 & 0.11 & 14
    & & 1.3$\times$10$^{14}$ & 1.7$\times$10$^{11}$
    & & 1.2$\times$10$^{14}$ & 2.5$\times$10$^{11}$ & & +3.4 \\ ~ \\
    G404 & 310 & 0.07 & 9
    & & 9.7$\times$10$^{13}$ & $<$1$\times$10$^{10}$
    & & 8.6$\times$10$^{13}$ & $<$1$\times$10$^{10}$ & & +1.0 \\
    G373 & 160 & 0.10 & 16
    & & 1.1$\times$10$^{14}$ & $<$1$\times$10$^{10}$
    & & 9.1$\times$10$^{13}$ & 0.4$\times$10$^{11}$ & & +1.1 \\
    G372 & 110 & 0.12 & 17
    & & 1.1$\times$10$^{14}$ & 2.2$\times$10$^{11}$
    & & 7.6$\times$10$^{13}$ & 2.6$\times$10$^{11}$ & & +1.1 \\
    G370 & ~\,75 & 0.18 & 16
    & & 2.3$\times$10$^{14}$ & 2.7$\times$10$^{11}$
    & & 1.9$\times$10$^{14}$ & 3.2$\times$10$^{11}$ & & +1.3 \\ ~ \\
    error\footnotemark[3] & $\pm$1 & $\pm$0.01 & $\pm$1
    & & $\pm$3\% & $\pm$3\% & & $\pm$3\% & $\pm$3\% & & $\pm$0.02
    \end{tabular}
    \end{ruledtabular}
    \footnotetext[1]{For all the ``G'' wafers, the number cited in this column is the average of measurements from two Hall bars.}
    \footnotetext[2]{This wafer is listed separately because it was grown in a different MBE chamber, part of another set described in Ref.\,\cite{Wendy10}.}
    \footnotetext[3]{These uncertainties apply to the whole column.}
    \label{tab:Fits}
\end{table*}

The gain/loss in mobility shown in Figure \ref{Fig6} is persistent at low temperatures, lasting for weeks \footnote{This was observed in sample H, measured before/after a university-wide shutdown due to COVID-19}. Furthermore, after a thermal cycle to room temperature and back to low temperatures, samples recover their dark transport characteristics, \textit{e.g.} as shown in Figure \ref{Fig4}(a). Sample E was illuminated during the first cooldown, cycled to room temperature, and cooled down again. Its transport characteristics in both cooldowns are nearly identical. In other words, like their modulation-doped cousins, dopant-free 2DEGs display the persistent photoconductivity effect.

We now turn to modeling to gain insight, using equation (\ref{eq:Matthiesen}) in conjunction with eqns.\,(\ref{eq:1byTauIRfinal})$-$(\ref{eq:1byTauBI-GaAsfinal}). Model parameters for each wafer, such as the 2DEG depth $d$, are listed in Table \ref{tab:Fits}. The parameters $d$, $\Delta$, and $\Lambda$ do not change between before and after illumination.

For all wafers, we set $N_{\textsc{bi-1}} = N_{\textsc{bi-2}} = \overline{N_{\textsc{bi}}}$, the average background impurity concentration. When fitting the mobilities of W639/W640/W641 (from Series I) before/after illumination, the same $\overline{N_{\textsc{bi}}}$ was imposed on all three wafers, and surface charge density $N_{\textsc{sc}}$ was used as the only unrestricted free variable. Best fits were obtained by minimizing the sum of squared differences between experiment and theory. Parameter values for the best fit are listed in Table \ref{tab:Fits}. For the Series II wafers (G370/G372/G373/G404), both $\overline{N_{\textsc{bi}}}$ and $N_{\textsc{sc}}$ were used as unrestricted free variables for fitting mobilities in each wafer, if $N_{\textsc{sc}}$ was not negligible. For example, for wafer G404, $N_{\textsc{sc}}$ was negligible and a one-parameter fit ($\overline{N_{\textsc{bi}}}$) was sufficient to model its mobility both before and after illumination. Parameter values for the best fit are listed in Table \ref{tab:Fits}.

The first common theme surmised from Table \ref{tab:Fits} to all 2DEGs from the ``W'' and ``G'' wafer series is that illumination appears to reduce the net average density of ionized/charged background impurities $\overline{N_{\textsc{bi}}}$, and is responsible for the improved mobilities (by up to +25\%, at  the same electron density) in the 2DEGs that are 75 nm or more away from the surface. One must therefore answer the question: what type of charged/ionized impurities can, upon illumination, be converted into neutral ones?

The usual candidate for the persistent photoconductivity effect in modulation-doped GaAs/AlGaAs 2DEGs are DX centers in AlGaAs \cite{Mooney90,Mooney91}, deep traps consisting of a single impurity atom and an associated crystal lattice deformation. They are often linked to Si impurities, the most common intentional \textit{n}\nobreakdash-type dopant in GaAs/AlGaAs 2DEGs, but can also arise from other impurity atoms such as Ge, Sn, Se, S, and Te \cite{Mooney90}. Of these, we will mostly focus on Si in the present discussion, the most studied and a likely background impurity in our MBE chambers due to the presence of Si effusion cells. We note however that sulfur is one of the two most common impurities present in high-purity arsenic sources \cite{Skromme85,Larkins87,Manfra14} (the other being carbon), and has been predicted to form DX$^-$ centers in both AlGaAs \cite{Mooney90} \textit{and} GaAs \cite{ParkCH96,DuMH05}.

Scenarios involving the conversion of negatively-charged DX$^-$ centers into neutral shallow donors ($d^0$) through illumination could potentially explain our observations. One such scenario could be DX$^- + ~d^+ \Rightarrow 2d^0$, as observed in modulation-doped GaAs/AlGaAs 2DEGs \cite{Hayne96,Hayne98}, in which two ionized impurities before illumination are converted into two neutral impurities after illumination by transferring an electron $e^-$ from the DX center to the ionized shallow donor $d^+$. The resulting reduced scattering from $\overline{N_{\textsc{bi}}}$ would increase the mobility at a given carrier concentration. Furthermore, since the bandstructure in our samples is essentially flat at $V_{\text{topgate}} = 0$, there is no energy barrier preventing free electrons in AlGaAs from migrating to GaAs. This is not the case in modulation-doped GaAs/AlGaAs heterostructures. Shallow donors in GaAs could therefore be available to receive electrons released by the neutralized DX centers in the AlGaAs. This would in turn reduce $N_{\textsc{bi-2}}$ (background impurities in GaAs), which would have more impact on mobility than neutralizing a shallow donor in AlGaAs.

However, the shallow neutral donor state of Si dopants is not stable, and is known to decay back to a positively-charged state $d^0 \rightarrow d^+ + \,e^-$ after some time  \cite{Mooney91,Hayne96}, ranging from seconds to tens of hours \footnote{In typical modulation-doped 2DEGs, the $d^0$ state's lifetime depends on the tunneling rate through the AlGaAs barrier separating the impurity from the 2DEG.}. It cannot revert back to the DX$^-$ state. Similarly, Silicon's neutral deep donor state DX$^0$ (obtained by DX$^- \Rightarrow ~$DX$^0 \, + \, e^-$ after illumination) is also metastable, and reverts back to either DX$^-$ (through thermal electron re-capture) or $d^+$ (through further photo-ionization) \cite{Mooney91}. This naturally leads one to ask the following two questions about the experiments of Figure \ref{Fig6}. Did metastable states play a major role in the mobility increase? On the other hand, were the measurements after illumination performed with all impurities in a new equilibrium state?

\begin{figure}[t]
    \includegraphics[width=\columnwidth]{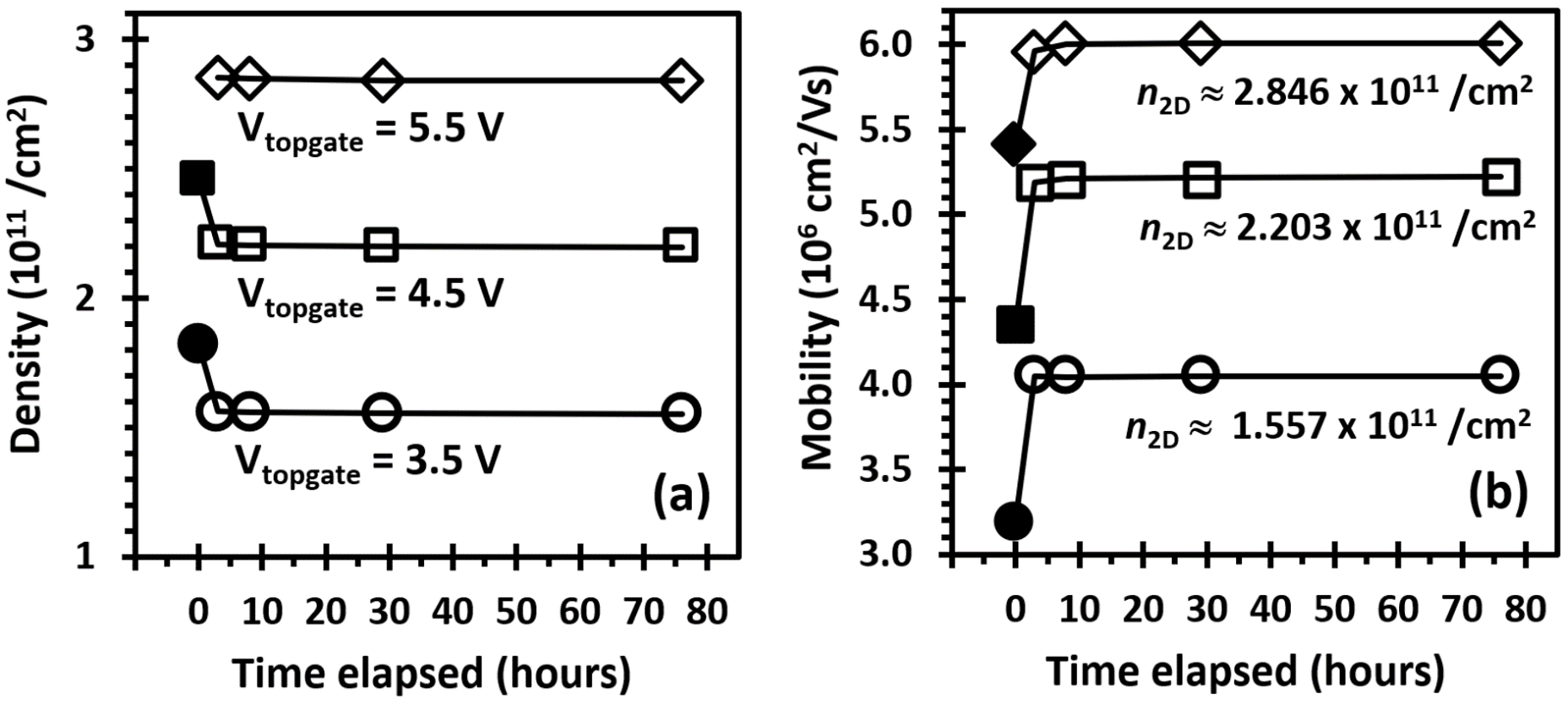}
    \caption{Stability with time of transport characteristics in sample J, after illumination performed at time $t=0$ hours. Between measurements, the topgate and the 2DEG were grounded. (a) Electron density versus time, at a fixed topgate voltage. For comparison, the densities before illumination are shown (closed symbols) at $t=0$, for $V_{\text{topgate}} \leqslant 5$~V. Error bars due to measurement uncertainty in the density ($\pm$0.03\%) are smaller than the symbols and are not shown.  (b) Mobility versus time, at the same fixed topgate voltages as in panel (a), with the corresponding approximate density indicated. For comparison, the mobilities before illumination are shown (closed symbols) at $t=0$, interpolated from the curve fit in Fig.\,\ref{Fig6}(g) at the same corresponding electron density. Error bars due to measurement uncertainty in the mobility ($\pm$0.2\%) are smaller than the symbols and are not shown. In both panels, lines are guides to the eye.}
    \label{FigPPC}
\end{figure}  

Figure \ref{FigPPC} answers those two questions: no and yes, respectively. Both the density [Fig.\,\ref{FigPPC}(a)] and mobility [Fig.\,\ref{FigPPC}(b)] are stable for more than 3 days after illumination, thus demonstrating persistent photoconductivity and confirming the sample is in the equilibrium regime. Had a large number of $d^0$ or DX$^0$ states decayed to a charged state ($d^+$ or DX$^-$) during that time, the mobility would have decreased \footnote{In fact, there is a very small increase (by +\,0.6\%) in the mobility over 3 days for $n_{\text{2D}} \approx 2.846 \times 10^{11}$ /cm$^2$ and $n_{\text{2D}} \approx 2.203 \times 10^{11}$ /cm$^2$ (but not for $n_{\text{2D}} \approx 1.557 \times 10^{11}$ /cm$^2$). This mobility increase appears real: it is not due to drift of the gating characteristics of the SiO$_2$ dielectric, because the density decreases (by $-$\,0.5\%) over the same 3-day period. All else being equal, one would expect mobility to decrease if the density decreases in this sample, see Fig.\,\ref{Fig6}(g). In any case, this small mobility increase pales in comparison to the large mobility increase (up to +\,30\%) immediately after illumination, confirming that metastable shallow donors play only a minor role in our observations.}, as is indeed observed in modulation-doped GaAs 2DEGs \cite{Hayne96}. Thus, neutral metastable states are not the primary cause for the $+$30\% mobility increase observed in sample J in Fig.\,\ref{Fig6}(g). We will assume this also applies to all samples from Fig.\,\ref{Fig6}.

The illumination scenarios discussed so far involving conventional DX centers will therefore collapse into the net ``reaction'' DX$^- \Rightarrow d^+ + \,2e^-$ (where the two released electrons $e^-$ are swept away by the ohmic contacts or the surface). This process essentially swaps a negatively-charged impurity (DX$^-$) for a positively-charged impurity ($d^+$). Recalling that the formation of DX centers follows the reaction $2d^0 \rightarrow ~$DX$^- +\, d^+$ \cite{Chadi89}, an ionized shallow donor $d^+$ must be present for every DX$^-$ center present in GaAs or AlGaAs, before illumination. Thus, in our undoped samples, the distribution of ionized background impurities changes from a mixture of DX$^-$ and $d^+$ states before illumination to a distribution of only $d^+$ states after illumination. This type of change in impurity distribution is known to increase mobility in modulation-doped 2DEGs (for a constant carrier density) \cite{Hayne98}, but is not well modeled by Eqns.\,(\ref{eq:1byTauBI-AlGaAsfinal}) and (\ref{eq:1byTauBI-GaAsfinal}), which instead reduce $\overline{N_{\textsc{bi}}}$ to account for the increase in mobility. Therefore, illumination causing the conversion of most DX$^-$ centers into $d^+$ states could be consistent with our experimental data.

Other scenarios involving exotic deep donor DX complexes or DX-like states cannot be ruled out, and could be consistent with our experimental data and modeling. For example, one could speculate that the long-lived neutral deep-donor DX$^0$-like state in GaAs \textit{after illumination} (DD$^0$), reported by Carey et al.\,\cite{Carey96}, could fulfill a similar role as the short-lived $d^0$ state in the processes/reactions discussed above, \textit{i.e.} DX$^- \Rightarrow$ DD$^0 + \,e^-$. Such a process taking place in GaAs, where the 2DEG resides, would have much more impact on mobility than in AlGaAs, relatively far away from the 2DEG.

Is there another type of impurity \textendash ~other than deep donor DX centers \textendash ~that would, upon illumination, transition from a charged state to a neutral state in both GaAs and AlGaAs, and cause a persistent photoconductivity effect? A. M. See \textit{et al.} \cite{See15} have proposed charge neutralization of ionized carbon impurities could fit these requirements. Carbon atoms are common background impurities in MBE chambers and are present in both AlGaAs and GaAs \cite{Skromme85,Larkins87,Manfra14}. Photoluminescence studies have identified deep acceptor states associated with carbon \cite{Skromme85}. Persistent photoconductivity has been reported after illumination of carbon modulation-doped GaAs/AlGaAs two-dimensional holes gases (2DHG) \cite{Manfra05,Gerl07,Watson12}. This strongly suggests the existence of deep acceptor states associated with a lattice deformation (negative-$U$ model), so-called ``AX centers''. These are predicted for carbon impurities in AlGaN alloys \cite{ParkCH97}, but could be possible in other III-V materials \footnote{The larger sizes of the As atom (linked to covalent bond lengths), of the lattice constant, and of the bandgap of (Ga,Al)As relative to AlGaN may favor the formation of a stable AX center in (Ga,Al)As.}, raising the possibility of the illumination reaction AX$^+ + \,a^- \Rightarrow \,2a^0$ \cite{ParkCH97}, where $a^-$ ($a^0$) is a charged (neutral) shallow acceptor. In our experiments, the two-step mechanism for charge neutralization would first involve the band-to-band optical excitation of an electron trapped in an $a^-$ impurity in either GaAs or AlGaAs, since the photon energy of the red LED ($\sim$\,2 eV) exceeds the bandgaps of both GaAs ($\sim$\,1.5 eV) and AlGaAs ($\sim$\,1.9 eV). Second, the liberated electron is captured by a nearby AX$^+$ center, in GaAs or AlGaAs. This process would thus convert two charged impurity states ($a^-$, AX$^+$) into two neutral ones ($a^0$), and increase mobility. For the above mechanism to be viable, the $a^0$ state after illumination must be stable in time. Another requirement for this mechanism's viability is that most carbon impurities must already be ionized before illumination. This is indeed the case: Giannini \textit{et al.} reported ionization rates of more than 80\% for carbon impurities in GaAs and AlGaAs \cite{Giannini93}, while ionization rates of up to 100\% have been reported if the carbon doping density is less than 3$\times$10$^{17}$ cm$^{-3}$ \cite{ItoH93}, the relevant regime in the samples presented here. Thus, charge neutralization of acceptor impurities after illumination could be consistent with our experimental data and modeling.

\begin{figure}[t]
    \includegraphics[width=1.0\columnwidth]{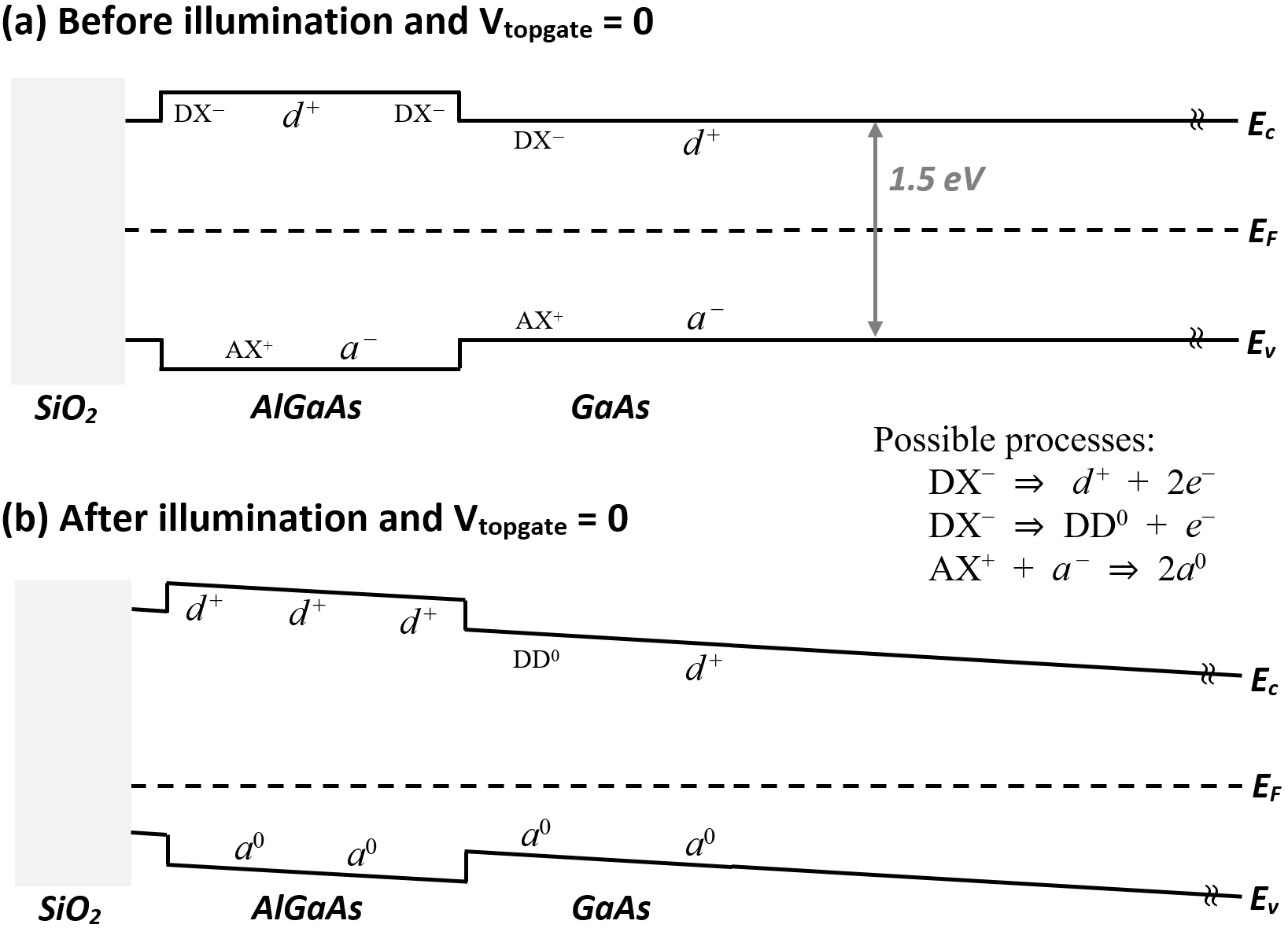}
    \caption{Schematic of the bandstructure of a hypothetical Hall bar from a Series I wafer [see Fig.\,1] with a SiO$_2$ gate dielectric at $V_{\text{topgate}}=0$ (not conducting): (a) before illumination, and (b) after illumination. Although the bandstructure has some curvature, it is at least 1,000$\times$ less pronounced than in intentionally modulation-doped 2DEGs. The gate dielectric and surface charge dominate the electric field [see section IV.B], and band bending due to charged background impurities is not visible at this scale. Processes discussed in the main text are shown.}
    \label{FigBandVtg0}
\end{figure}

The second common theme surmised from Table \ref{tab:Fits} is that illumination appears to \textit{increase} the surface charge density. This could be caused by the activation of surface states/traps by light. Another possible cause is the accumulation of electrons at the surface for the sample to maintain overall charge neutrality, because of electrons released by impurities (such as DX centers) or band-to-band photo-excited electrons. In 7 out of 9 ambipolar samples from Series II where transport properties of 2DHGs were also measured, the observed change before/after illumination in threshold voltage $\Delta V_{\text{th}}$ was the \textit{same} for both the 2DEG (electrons) and the 2DHG (holes) in the same Hall bar; in all cases without exception, a higher hole density was reached for the same topgate voltage after illumination than before. The increase in surface charge density is larger for the shallower 2DEGs, and this is reflected in both the ``W'' and ``G'' wafers by the increasing change in $\Delta V_{\text{th}}$ as the 2DEG depth $d$ becomes smaller. This suggests a net negative surface charge. Here, the magnitude of $\Delta V_{\text{th}}$ is affected by the nature of the gate dielectric: $\Delta V_{\text{th}}$ is much larger with polyimide than with SiO$_2$ (also see next section IV.B, ``biased illumination''). In their study of illumination in HIGFET devices, Fujita \textit{et al.} \cite{FujitaT21} observed $V_{\text{th}}$ decreased ($\Delta V_{\text{th}}<0$) after a single, long illumination at a wavelength of 780 nm ($\sim$1.6 eV, less than the AlGaAs bandgap). This is the opposite trend to what we observe. However, they also inferred the appearance after illumination of a large population of holes in their GaAs substrate: this could account for their observation of a decrease in $V_{\text{th}}$. Indeed, in another experiment with multiple low-intensity light pulses, after a large initial drop in $V_{\text{th}}$ (possibly mostly due to the positively-charged substrate), they observed subsequent small increases in $V_{\text{th}}$ with increasing illumination \cite{FujitaT21}, consistent with the data shown here. Figure \ref{FigBandVtg0} shows bandstructure schematics before and after an unbiased illumination, summarizing the possible scenarios mentioned in this section.

Finally, for wafer V627 [Figure \ref{Fig6}(d)], modeling is treated separately from the others. It is the shallowest 2DEG of the dataset presented here, located only 30 nm below the surface. The composition of its Al$_x$Ga$_{1-x}$As barrier layer is also different from all other wafers ($x=0.90$ instead of $x=0.33$). After illumination \footnote{Only one Hall bar from wafer was measured after illumination}, it suffers from a dramatic loss in mobility (by 50$-$85\%), presumably resulting from the increased scattering associated with surface charges. Data after illumination could not be fit to Boltzmann transport equations (\ref{eq:1byTauIRfinal})$-$(\ref{eq:1byTauBI-GaAsfinal}). However, it could instead be fit [red dashed line in Fig.\,\ref{Fig6}(d)] to the equation:
\begin{equation}
\mu = A_0 (n_{\textsc{2d}} - n_c)^{4/3}
\label{eq:percolation}
\end{equation}
with $A_0 = 1.47 \times 10^{-2}$ cm$^{2/3}$/Vs and critical density $n_c = 3.9 \times 10^{10}$ /cm$^2$ as fit parameters. Equation (\ref{eq:percolation}) describes transport in the regime near the 2D percolation threshold \cite{Palevski84, PercolationBook-1, PercolationBook-2}, when the 2DEG breaks up in ``puddles'' and ceases to be continuous. Before illumination, data from the upper mobility range can be fit to equations (\ref{eq:1byTauIRfinal})$-$(\ref{eq:1byTauBI-GaAsfinal}) with parameters listed in Table \ref{tab:Fits} [black solid line in Fig.\,\ref{Fig6}(d)], and data from the lower mobility range can be fit to equation (\ref{eq:percolation}) [black dashed line in Fig.\,\ref{Fig6}(d)] with $A_0 = 2.89 \times 10^{-2}$ cm$^{2/3}$/Vs and $n_c = 1.9 \times 10^{10}$ /cm$^2$. The critical density $n_c$ is higher after illumination than that before illumination, consistent with the observed decrease in mobility due to the corresponding increase in disorder.

\subsection{Biased illumination, while $V_{\text{topgate}}$\,$\neq$\,0}
\label{section:Vtg-neq0}

Since polyimide can leak when illuminated while $V_{\text{topgate}}$\,$\neq$\,0, biased illumination was only performed on Hall bars from Series II, with a SiO$_2$ gate dielectric. Devices were cooled down in the dark, and illuminated at $T$\,=\,1.4 K. In order to separate the effects of unbiased from biased illuminations, devices were initially illuminated for 6 minutes while keeping  $V_{\text{topgate}} = 0$. Heat dissipation from the LED caused a nominal temperature increase of $T<1.8$ K. After this initial illumination, subsequent biased illuminations were carried out by illuminating for one minute with the topgate held at finite voltage values. After the LED was turned off, the topgate was set to zero voltage, and the sample cooled back down to $T$\,=\,1.4 K before measurements would begin. Thus, most biased illuminations on a particular device were performed during a single cooldown.

\begin{figure}[t]
    \includegraphics[width=\columnwidth]{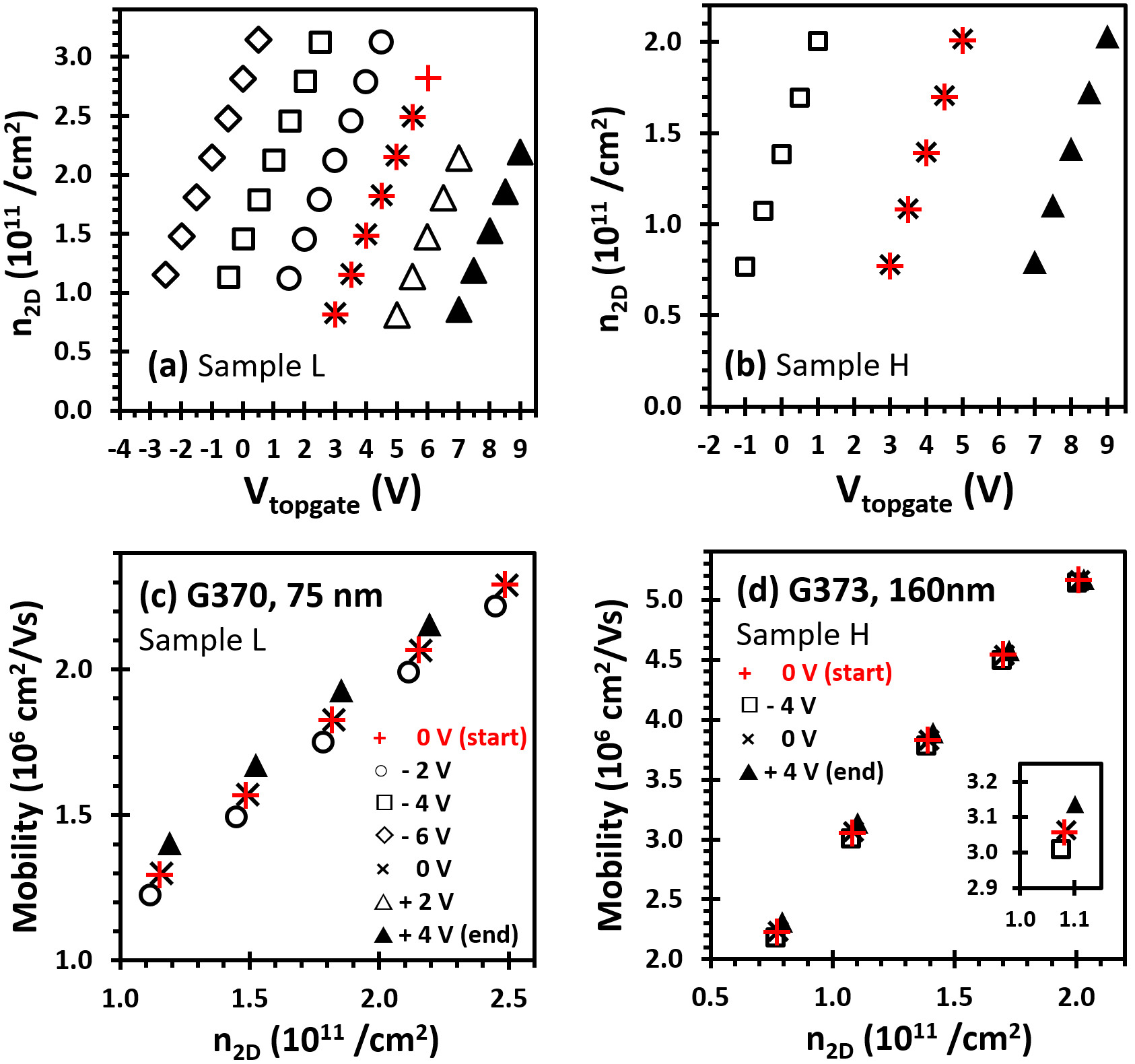}
    \caption{(color online) Electron density versus topgate voltage after multiple biased illuminations on: (a) shallow wafer G370, and (b) deep wafer G373. Symbols in all four panels
    are defined in panel (c), and are in the order of biased illuminations performed during the same cooldown, with `start' being the first. In all cases, the electron density versus topgate voltage relation is shifted by the voltage at which the biased illumination was performed. Note how the characteristics of the initial `0V' biased illumination (red `+' symbols) are recovered when another biased illumination at $V_{\text{topgate}}$ = 0 V is performed (black `$\times$' symbols) after three biased illuminations are performed at $V_{\text{topgate}}$ = $-2$~V, $-4$~V, and $-6$~V. Electron mobilities after multiple biased illuminations are shown for a: (c) shallow wafer G370, and (d) deep wafer G373. For clarity, only a selection of biased illuminations are shown in these panels. The inset in panel (d) has the same axes and units as in the main figure; it is a magnified view of the data points near $n_{\textsc{2d}} = 1.1 \times 10^{11}$ /cm$^2$. }
    \label{FigNEQ0}
\end{figure}  

Akin to biased cooling \cite{Buks94-A, Buks94-B, Coleridge97, Pioro05-B}, the density-topgate voltage functions $n_{\textsc{2d}}(V_{\text{topgate}})$ in Figures \ref{FigNEQ0}(a) and \ref{FigNEQ0}(b) are shifted by the voltage at which the topgate was held during illumination. For each device, the slopes of all $n_{\textsc{2d}}(V_{\text{topgate}})$ are the same as each other and the same as that for illumination at $V_{\text{topgate}}$\,=\,0; the topgate-2DEG capacitance does not change. Remarkably, biased illumination appears to be a \textit{reversible process}, relative to illumination at $V_{\text{topgate}}$\,=\,0. This is illustrated in both Figures \ref{FigNEQ0}(a) and \ref{FigNEQ0}(b): after an initial illumination at $V_{\text{topgate}}$\,=\,0 (red `+' symbols), a series of biased illuminations are performed before repeating an illumination at $V_{\text{topgate}}$\,=\,0 (black `$\times$' symbols). The $n_{\textsc{2d}}(V_{\text{topgate}})$ function of the initial $V_{\text{topgate}}$\,=\,0 illumination is recovered after the subsequent $V_{\text{topgate}}$\,=\,0 illumination (the `+' and `$\times$' symbols line up almost perfectly). Recovery of original characteristics is not limited only to the $V_{\text{topgate}}$\,=\,0 illumination, as we have confirmed that $n_{\textsc{2d}}(V_{\text{topgate}})$ of \textit{any} biased illumination at $V_{\text{topgate}}$\,$=$\,$V_0$ can be recovered by illuminating again with the same topgate voltage $V_0$.

Although the effects of biased illumination on mobility are small (with differences of up to 7\% between the smallest and largest mobilities), these are still larger than measurement uncertainties ($<$0.3\%). Figures \ref{FigNEQ0}(c) and \ref{FigNEQ0}(d) show transport measurements on two 2DEGs, located 75 nm and 160 nm below the wafer surface respectively. One observation is that, for both 2DEG depths, biased illuminations performed when $V_{\text{topgate}} \geqslant +2$~V increase mobility, whereas those performed when $V_{\text{topgate}} \leqslant -2$~V decrease mobility. A second observation is that the mobility changes are larger in the shallower 2DEG than in the deeper 2DEG. A third observation is that mobility changes are \textit{reversible} for both shallow and deep 2DEGs, within the \textit{same} cooldown.

In both Figures \ref{FigNEQ0}(c) and \ref{FigNEQ0}(d), after an initial illumination at $V_{\text{topgate}}$\,=\,0 (red `+' symbols), a series of biased illuminations are performed before repeating an illumination at $V_{\text{topgate}}$\,=\,0 (black `$\times$' symbols). The mobility of the initial $V_{\text{topgate}}$\,=\,0 illumination is recovered after a subsequent $V_{\text{topgate}}$\,=\,0 illumination: the `+' and `$\times$' symbols line up almost perfectly. The mobility gain/loss must be accounted for by a decrease/increase in electron scattering.

\begin{figure}[t]
    \includegraphics[width=\columnwidth]{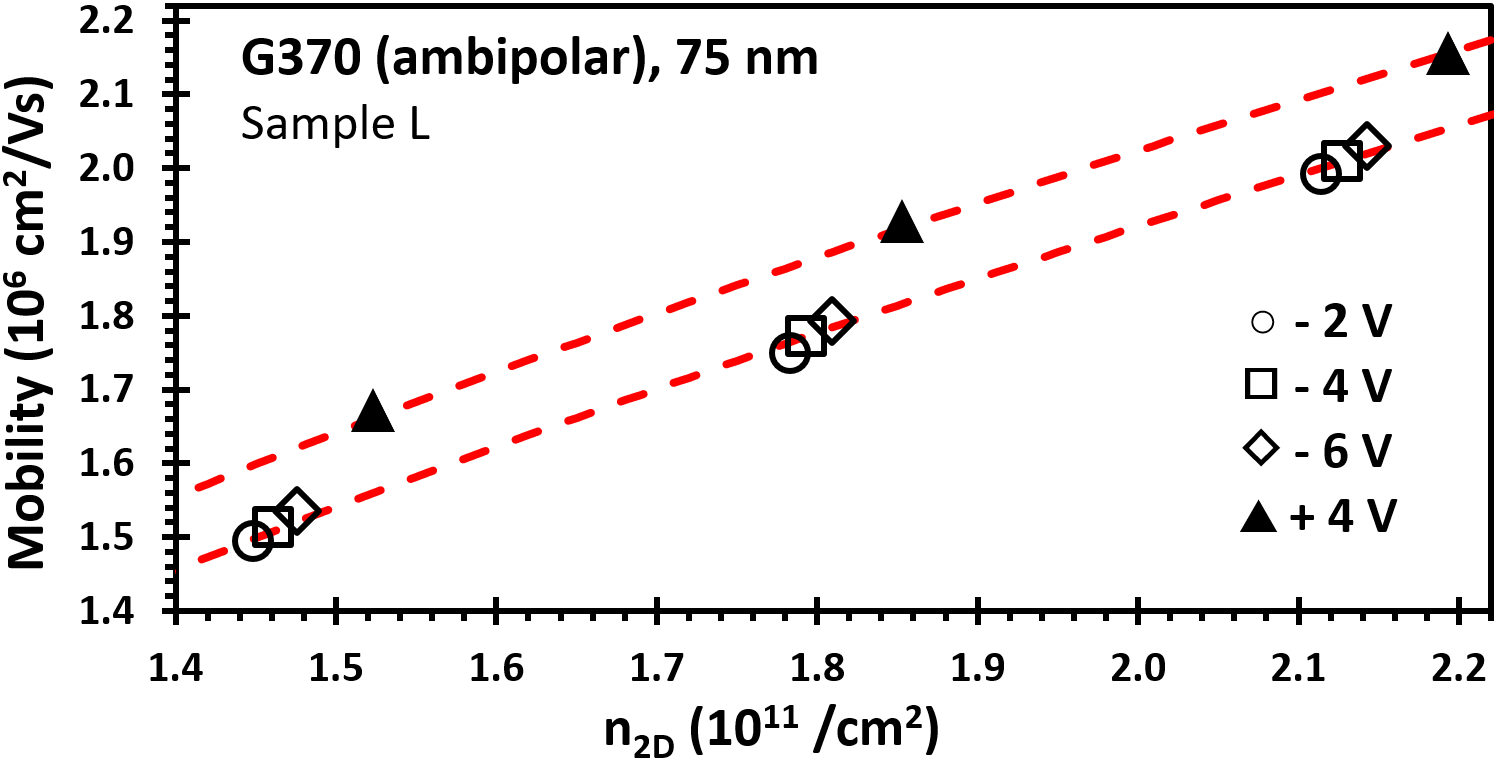}
    \caption{(color online) Electron mobilities for ambipolar sample L (with both \textit{n}-type and \textit{p}-type ohmic contacts) shown in Fig.\,\ref{FigNEQ0}(c), and close-up view of the multiple biased illuminations at $V_{\text{topgate}}$ = $-2$, $-4$, $-6$, and $+4$ V in chronological order. The two dashed lines are otherwise identical mobility simulations, except for a difference of $\Delta N_{\textsc{sc}} = 8 \times 10^{10}$ /cm$^2$ in surface charge density, using the Boltzmann transport model described in section \ref{section:theory}. The two key features to note here are: (i) all the $V_{\text{topgate}} < 0$ biased illumination mobilities fall on the same mobility curve $\mu(n_{\textsc{2d}})$, and (ii) the mobility after the $V_{\text{topgate}}$ = $+4$ V biased illumination increases.}
    \label{FigAmbi}
\end{figure}  

A possible culprit could be the gate dielectric. The amorphous SiO$_2$ layer contains a very large number of defects (relative to single crystal GaAs/AlGaAs), a fraction of which could populate or depopulate with electrons during biased illumination, in response to the finite topgate voltage. This certainly could explain the voltage shifts (equal to the topgate voltage value during biased illumination) in $n_{\textsc{2d}}(V_{\text{topgate}})$ observed in Figures \ref{FigNEQ0}(a) and \ref{FigNEQ0}(b). This scenario would also be consistent with the deeper 2DEGs experiencing smaller mobility gains/losses, since they are further away from the SiO$_2$ layer. Before illumination, the range of topgate voltages without hysteresis is restricted to approximately $|V_{\text{topgate}}|$\,$\lesssim$\,5 Volts. However, after illumination (whether at $V_{\text{topgate}}$\,=\,0 or $V_{\text{topgate}}$\,$\neq$\,0), the range of topgate voltages without hysteresis is extended to $|V_{\text{topgate}}|$\,$\lesssim$\,9 Volts. This suggests illumination does introduce some changes to the SiO$_2$ layer (ionization of defects), and is consistent with the scenario depicted above.

However, Figure \ref{FigAmbi} presents a puzzle that cannot be explained by the scenario above. Upon close inspection, the mobilities after biased illuminations at $V_{\text{topgate}} =$ $-2$~V, $-4$~V, and $-6$~V on sample L (wafer G370) all lie nearly on the same mobility curve $\mu(n_{\textsc{2d}})$. In other words, the mobility loss has saturated after the $V_{\text{topgate}} =$ $-2$~V biased illumination, which implies that the number of scattering centers in the SiO$_2$ layer is no longer increasing with biased illuminations at more negative topgate voltages. Yet, the $n_{\textsc{2d}}(V_{\text{topgate}})$ relation in Figure \ref{FigNEQ0}(a) shows no signs of saturation, with ever larger topgate voltage shifts. The latter implies an increasing number of active defects in the SiO$_2$ layer from biased illuminations with increasing topgate voltages. Both statements cannot be simultaneously true.

To resolve the inconsistency outlined above, we propose that the behavior of the relation  $n_{\textsc{2d}}(V_{\text{topgate}})$ is primarily affected by defect-driven charging effects in the SiO$_2$ layer, and that the behavior of the relation $\mu(n_{\textsc{2d}})$ is primarily affected by changes in surface charge density $N_{\textsc{sc}}$. In this new scenario, mobility increases (decreases) when the surface charge density decreases (increases) due to $V_{\text{topgate}}$\,$>$\,0 ($V_{\text{topgate}}$\,$<$\,0) biased illuminations. One possible mechanism for a gain in mobility is the (re-)capture of electrons by charged surface defects, facilitated by $V_{\text{topgate}}$\,$>$\,0. The loss in mobility when $V_{\text{topgate}}$\,$<$\,0 would correspond to further ionization of ``dangling'' bonds at the surface, i.e. the GaAs/SiO$_2$ interface. The saturation of mobility loss occurs when all available surface defects have been ionized.

\begin{figure}[t]
    \includegraphics[width=\columnwidth]{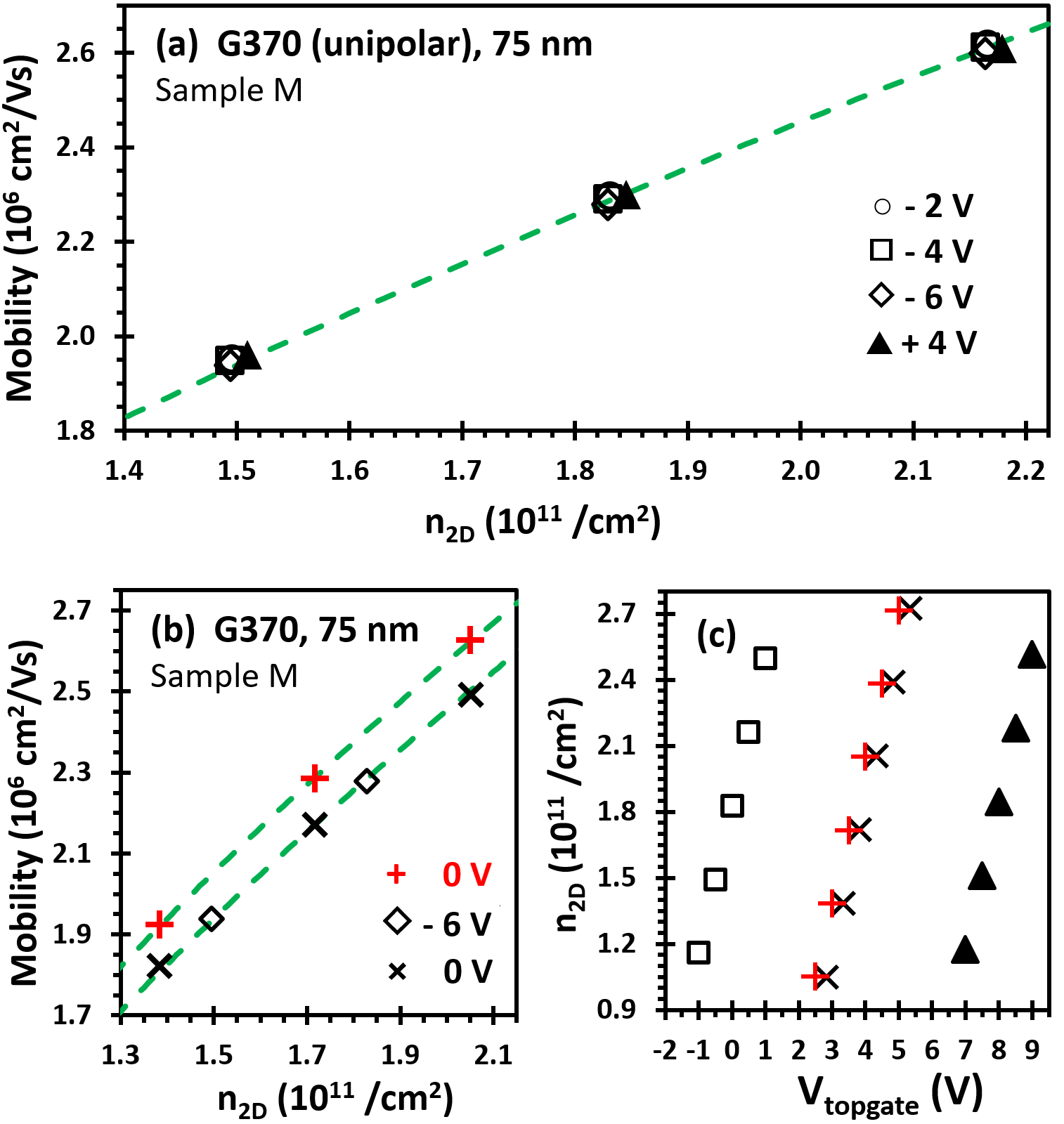}
    \caption{(color online) Non-reversibility of transport characteristics after biased illuminations on unipolar sample M (with only \textit{n}-type ohmic contacts). For all panels, the chronological order of biased illuminations and symbols used are identical to that indicated in Figure \ref{FigNEQ0}(c). (a) Unlike ambipolar Hall bars [Fig.\,\ref{FigAmbi}], the mobilities after the $V_{\text{topgate}} = +4$ V biased illumination (`$\blacktriangle$') do not increase, but remain on the same mobility curve $\mu(n_{\textsc{2d}})$ (dashed line) as those from $V_{\text{topgate}} < 0$ biased illuminations (open symbols). (b) Likewise, the mobilities of the initial $V_{\text{topgate}}$ = 0 V biased illumination (`+') are not recovered when another $V_{\text{topgate}}$ = 0 V biased illumination is performed (`$\times$') after the $V_{\text{topgate}} < 0$ biased illuminations (for clarity, only the `$\diamond$' symbols are shown), in contrast to behavior shown in Figs.\,\ref{FigNEQ0} and \ref{FigAmbi} for ambipolar devices. The two dashed lines are otherwise identical $\mu(n_{\textsc{2d}})$ mobility simulations, except for a difference of $\Delta N_{\textsc{sc}} = 6 \times 10^{10}$ /cm$^2$ in surface charge density, using the Boltzmann transport model described in section \ref{section:theory}. (c) Unlike ambipolar Hall bars [see Figures \ref{FigNEQ0}(a) and \ref{FigNEQ0}(b)], the $n_{\textsc{2d}}(V_{\text{topgate}})$ curve of the original $V_{\text{topgate}} = 0$ biased illumination (`+') is not recovered after the subsequent $V_{\text{topgate}} = 0$ biased illumination (`$\times$'). The +\,0.33 V shift to a higher threshold topgate voltage to reach the same electron density is consistent with our prediction of a permanent increase in surface charge density $N_{\textsc{sc}}$ after a $V_{\text{topgate}}<0$ biased illumination (see main text).}
    \label{FigUni}
\end{figure}  

Although there are far fewer available defects at the surface of single-crystal GaAs than in a 300 nm-thick amorphous SiO$_2$ layer, ionized impurities at the wafer surface are much more effective at scattering electrons: (i) they are physically much closer to the 2DEG, and, in the parlance used for quantum dot transport, (ii) they have a much bigger lever arm because of the higher relative dielectric constant in Al$_{0.3}$Ga$_{0.7}$As ($\epsilon_r$\,$\approx$\,12) relative to our PECVD SiO$_2$ ($\epsilon_r$\,$\approx$\,3.5). Recalling eqn.\,(\ref{eq:1byTauSCfinal}), the scattering rate of electrons in a 2DEG due to an ionized impurity is an exponentially decreasing function of distance. So, even if there are far more SiO$_2$ bulk defects than surface states (i.e. GaAs/SiO$_2$ interface states), the latter are exponentially more effective at increasing/decreasing the 2DEG mobility.

This new scenario is consistent with both mobility loss saturation (Fig.\,\ref{FigAmbi}) and the decreasing effects of biased illumination with increasing 2DEG depth (Fig.\,\ref{FigNEQ0}). Next, we perform a sanity check on our proposed scenario. The mechanism for mobility gain after a $V_{\text{topgate}}$\,$>$\,0 biased illumination explicitly relies on electron-hole photo-generation, the (re\nobreakdash-)capture of photo-generated electrons by surface charge defects, and the presence of \textit{p}\nobreakdash-type ohmic contacts to sweep away the photo-generated holes. What if a sample does not have \textit{p}\nobreakdash-type ohmic contacts?

In that case, photo-generated holes would not be swept away by the \textit{p}\nobreakdash-type ohmic contacts, and would instead recombine with any available electrons, most likely the photo-generated electrons. The latter would thus not be available to be re-captured by ionized charge surface defects (\textit{i.e.}, $N_{\textsc{sc}}$ cannot decrease), and mobility would not increase any further. Figure \ref{FigUni} confirms that this is exactly what is observed in experiments on sample M, which has only \textit{n}\nobreakdash-type ohmic contacts but is otherwise identical in all other respects to samples from Series II. After an initial unbiased illumination (red `+' symbols in Fig.\,\ref{FigUni}), the mobility increases by 25\% from mobilities in the dark (not shown). This mobility gain does not require the presence of \textit{p}\nobreakdash-type ohmics. Next, a series of biased illuminations with $V_{\text{topgate}} < 0$ (`{\large$\circ$}', {`\scriptsize $\square$}', and `{\large$\diamond$}' symbols) are carried out, with mobility loss due to the increase in surface charge density (the ionization of all remaining charge surface defects) in Figs.\,\ref{FigUni}(a) and \ref{FigUni}(b). This also does not require the presence of \textit{p}\nobreakdash-type ohmic contacts. All mobilities fall onto the same $\mu(n_{\textsc{2d}})$ curve [Fig.\,\ref{FigUni}(a)]. Lastly, two more biased illuminations are carried out, one at $V_{\text{topgate}} = 0$ (black `$\times$' symbols) and one at $V_{\text{topgate}} = +4$~V (`$\blacktriangle$' symbols). Unlike the ambipolar Hall bars in Figure \ref{FigNEQ0}, the mobility of the unipolar Hall bar does not recover/increase [Fig.\,\ref{FigUni}(a)], but remains on the same $\mu(n_{\textsc{2d}})$ curve as that of the $V_{\text{topgate}} < 0$ biased illuminations [Fig.\,\ref{FigUni}(b)], as predicted by our proposed scenario. Finally, Fig.\,\ref{FigUni}(c) provides one last piece of evidence in favor of our scenario. After the $V_{\text{topgate}} < 0$ bias illuminations, the resulting `permanent' (for this cooldown) increase in surface charge density $N_{\textsc{sc}}$ causes a +0.33~V shift in $n_{\textsc{2d}}(V_{\text{topgate}})$ for the second $V_{\text{topgate}} = 0$ illumination (black `$\times$' symbols), relative to the first $V_{\text{topgate}} = 0$ illumination (red `$+$' symbols), to achieve the same electron density $n_{\textsc{2d}}$. This is in direct contrast to behavior observed in ambipolar devices [Figs.\,\ref{FigNEQ0}(c) and \ref{FigNEQ0}(d)], where bias illuminations are fully reversible.

\section{Discussion and Conclusion}

The main result of section \ref{section:Vtg-zero} indicates that unbiased illuminations can reduce scattering from ionized impurities concentration by up to 30\% in dopant-free 2DEGs, for the same electron density. The exact mechanism(s) through which this occurs remain unclear. The coventional mechanism involving Si impurity DX centers could perhaps explain the increase in mobility at the same carrier density. Other types of donor DX or DX-like centers could not be ruled out, whether with silicon or other impurity species (e.g. sulfur). Speculation about acceptor AX centers from carbon impurities in GaAs/AlGaAs led to a mechanism that could be consistent with data shown, provided photo-neutralized shallow acceptor states ($a^0$) are stable at low temperature. Whichever mechanism underpins our observations, the observed increase in mobility would most likely also occur in high-mobility \textit{n}\nobreakdash-type modulation-doped 2DEGs and  \textit{p}\nobreakdash-type modulation-doped 2DHGs.

Unbiased illumination is commonly used in fractional quantum Hall effect (FQHE) experiments in modulation-doped 2DEGs to improve the activation energies of FQHE states. Both the electron density and mobility increase significantly via the persistent photoconductivity effect. Interestingly, Samani et al. \cite{Samani14} were able to use a two-step unbiased illumination protocol that drastically improved their samples' FQHE characteristics, while retaining the same electron density and mobility as before illumination. Their samples use a delta-doping scheme that places Si dopants in GaAs ``doping'' wells \cite{Umansky09,Manfra14}, thus preventing the formation of DX centers. The improvement of their FQHE states after illumination is attributed to enhanced screening from dopant-dopant correlations \cite{FuX18,Buks94-B}, in a manner very reminiscent of overdoping \cite{Umansky09}. Could dopant-dopant correlations occur in dopant-free 2DEGs? This appears very unlikely. The Si delta-doping sheet density in the modulation-doped 2DEGs mentioned above is $(1-2)\times 10^{12}$ cm$^{-2}$, which corresponds to an average dopant-dopant separation of $8-11$ nm. With such proximity, strong correlations between dopants are indeed likely. In dopant-free 2DEGs, the background impurity volume density is of order $(1-2)\times 10^{14}$ cm$^{-3}$ or less, which corresponds to an impurity-impurity average separation of $450-570$ nm. Dopant-dopant correlations (and therefore screening) are unlikely to play a major role in the experiments described in this paper. Nevertheless, it would be interesting to observe the effects of illumination on FQHE states in a dopant-free 2DEG.

After an initial unbiased illumination, the effects of additional, biased illuminations (section \ref{section:Vtg-neq0}) on mobility $\mu(n_{\textsc{2d}})$ are very limited, with changes of $\sim$\,5\% at most for a narrow range of gate voltages. Outside of that gate voltage range, $\mu(n_{\textsc{2d}})$ no longer responds to biased illuminations, for either gain or \textendash ~unusually \textendash ~loss. The saturation of mobility losses is attributed to the activation of all available defects/states at the interface between the GaAs cap layer and the SiO$_2$ gate dielectric (the ``surface''). Whether these states/defects are located on the gate dielectric or the semiconductor side of this interface cannot be distinguished. While $\mu(n_{\textsc{2d}})$ no longer responds to biased illuminations, $n_{\textsc{2d}}(V_{\text{topgate}})$ still does, with the change in 2DEG turn-on threshold voltage $\Delta V_{\text{th}}$ exactly matching the $V_{\text{topgate}}$ at which the biased illumination was performed. In that regard only (ignoring the saturation of mobility losses), biased illumination bears some resemblance with bias cooling \cite{Pioro05-B}, in that it can ``lock in'' a built-in electric potential. However, unlike bias cooling, this built-in electric potential can be arbitrarily tuned \textit{in-situ} during the \textit{same} cooldown (with additional biased illuminations), and most likely originates in the gate dielectric.

In conclusion, we have shown that unbiased ($V_{\text{topgate}}$\,$=$\,0) and biased ($V_{\text{topgate}}$\,$\neq$\,0) illuminations have different effects on dopant-free 2DEGs, and presented possible mechanisms explaining the observed behavior. Unbiased illuminations increase (decrease) the mobility at the same electron density if the 2DEG depth below the surface is more (less) than $\sim$\,70 nm. Whether mobility increases or decreases results from the interplay between the reduction of charged background ionized impurities ($N_{\textsc{bi}}$) and the increase in surface charge density ($N_{\textsc{sc}}$) after illumination. Biased illuminations increase (decrease) mobilities, regardless of 2DEG depth, if the topgate voltage is $V_{\text{topgate}}$\,$>$\,0 ($V_{\text{topgate}}$\,$<$\,0), and is primarily driven by changes in the surface charge density ($N_{\textsc{sc}}$). The magnitude of the mobility gain/loss is larger (smaller) for 2DEGs that are close to (far from) the wafer surface. Remarkably, the effects of any specific biased illumination are fully reversible, both in mobility and 2DEG turn-on threshold voltages.

\begin{acknowledgments}
A.S., F.S., and W.Y.M. contributed equally to this work. The authors thank Christine Nicoll for illuminating discussions. I.F. thanks Toshiba Research Europe for financial support. This research was undertaken thanks in part to funding from the Canada First Research Excellence Fund (Transformative Quantum Technologies), Defence Research and Development Canada (DRDC), Canada's National Research Council (NRC) under contract W943741, Canada's Natural Sciences and Engineering Research Council (NSERC), as well as the UK's Engineering and Physical Research Council grants EP/K004077/1 and EP/J003417/1. The University of Waterloo's QNFCF facility was used for this work. This infrastructure would not be possible without the significant contributions of CFREF-TQT, CFI, ISED, the Ontario Ministry of Research and Innovation, and Mike and Ophelia Lazaridis. Their support is gratefully acknowledged.
\end{acknowledgments}

\appendix

\section{Interface Roughness}
\label{Appndx:IR}

The GaAs/AlGaAs interface where the 2DEG resides is not a perfectly smooth planar boundary, but consists instead of a textured plane where the electric field/barrier height have discontinuities in the plane. This interface roughness is characterized by the height $\Delta$ of irregularities (be it crystal defects, atomic steps, or other) in the $z$ direction and the separation distance $\Lambda$ between these irregularities in the \textbf{r} direction [see Fig.\,\ref{Fig3}]. The distribution of heights $\Delta(r)$ along the interface is assumed to be Gaussian: $<\Delta(r)\Delta(r')>=\Delta^2e^{-(r-r')^2/\Lambda^2}$. The effects of interface roughness are more pronounced at higher carrier densities: as the carrier density increases, the electron wavefunction increasingly overlaps with the interface, and thus scattering increases.  The wafer surface is used as a proxy for the GaAs/AlGaAs interface, and is characterized using an atomic force microscope (AFM), extracting values for $\Delta$ and $\Lambda$.

The scattering potential $|U(q)|^2$ for interface roughness is \cite{Gold88}:
\begin{equation}
|U(q)|^2_{\textsc{ir}} = \pi(\Lambda\Delta)^2
\left(\frac{e^2(\frac{1}{2}n_{\textsc{2d}}+N_{depl})}{\epsilon_0\epsilon_r}\right)^2
e^{-q^2\Lambda^2/4}
\end{equation}
\noindent which, after inserting the above expression in equation (\ref{eq:FermiGoldenRule}) and rewriting the integral in terms of $d\theta$, gives \cite{Ando77}:
\begin{eqnarray}
\frac{1}{\tau_{\textsc{ir}}}=\frac{m^*(\Lambda\Delta)^2}{2\hbar^3 k_F^2}\int_0^\pi \frac{q^2~\Gamma(q)^2}{\epsilon(q)^2~e^{\,q^2\Lambda^2/4}}~d\theta\label{eq:1byTauIR}\\
\text{with}\quad \Gamma(q)=\frac{e^2}{\epsilon_0\epsilon_r}\Big(\frac{n_{\textsc{2d}}}{2}+N_{depl}\Big)
\approx \frac{n_{\textsc{2d}}e^2}{2\epsilon_0\epsilon_r}  \label{eq:Gammaq}\\
N_{depl}=\sqrt{2\epsilon_0\epsilon_rN_aE_g}\qquad~~\,
\end{eqnarray}
\noindent where $N_{depl}$ is the depletion charge density, $N_a$ is the acceptor concentration in the GaAs layer, and $E_g$ is the bandgap in GaAs.

The depletion charge term arises if the material hosting the 2DEG is lightly p doped by impurity atoms ($N_{\textsc{bi-2}}$ in GaAs for example) or implantation (as is often the case in Si-based devices). This changes the overall bandstructure and affects the position of the electron wavefunction. However, because $\epsilon^{\text{GaAs}}_{r} \approx \epsilon^{\text{AlGaAs}}_r$ is assumed, the term $\Gamma(q)$ no longer depends on $\theta$ via the $\sin \frac{\theta}{2}$ term in $q$ and it can be pulled out of the $d\theta$ integral. Furthermore, since the background impurity concentration ($N_{\textsc{bi-2}}$) is much less than the carrier density ($n_{\textsc{2d}}$) in the experiments studied in this paper, \textit{i.e.} $N_{\textsc{bi-2}} \ll \frac{1}{2}n_{\textsc{2d}}$, we approximate $N_{depl}$\,$\approx$\,0 so that $\Gamma(q)=\frac{n_{\textsc{2d}}e^2}{2\epsilon_0\epsilon_r}$ in equation (\ref{eq:Gammaq}) and equation (\ref{eq:1byTauIRfinal}).

\section{Surface Charge}
\label{Appndx:SurfaceCharge}

\begin{figure}[t]
    \includegraphics[width=\columnwidth]{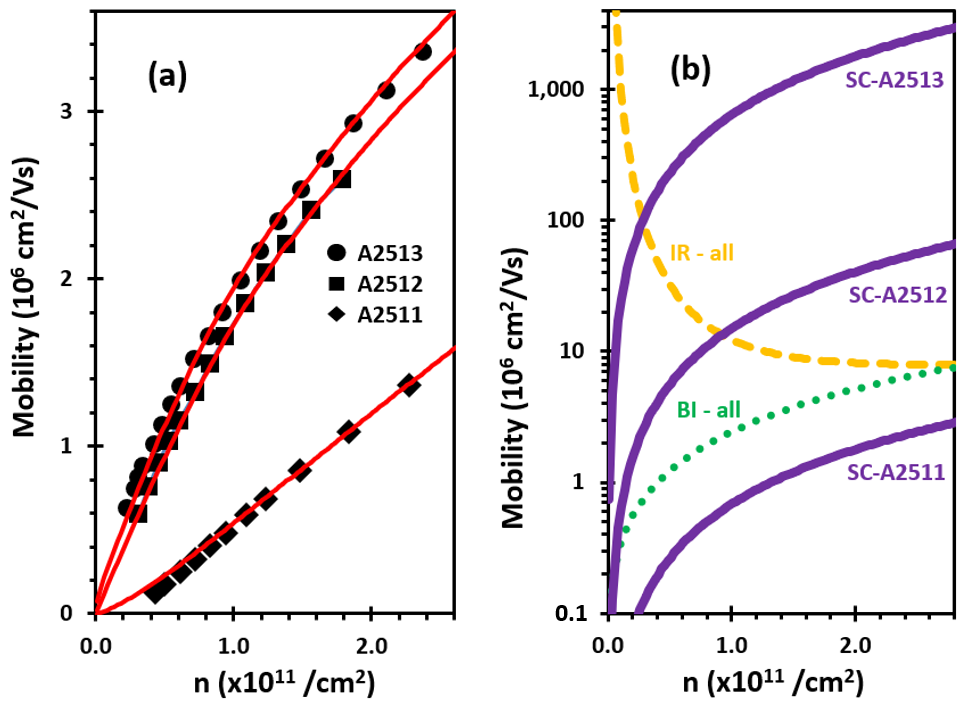}
    \caption{(color online) Effect of surface charge. (a) Experimental 2DEG mobilities at $T$\,=\,1.5\,K for 3 wafers (symbols) at different depths (taken from Ref.\,\onlinecite{WendyPhD}). Solid (red) lines are fits from the model in section III with the parameters listed in Table \ref{tab:Appndx-SCmobility}. The \textit{only} fit parameter that changes between the three red curves is the surface charge density $N_{\textsc{sc}}$. All data taken in the dark. (b) Breakdown of the contributions to the mobilities shown in panel (a). The (orange) dashed line is the contribution from interface roughness (IR), common to all 3 wafers. The (green) dotted line is the contribution from the average background impurity concentration (BI), common to all 3 wafers. The (purple) solid lines are the contributions from surface charge (SC) for each wafer. For wafer A2511 (40 nm deep 2DEG), surface charge is the most significant scattering mechanism over the entire electron density range, whereas surface charge is negligible for the 310 nm deep 2DEG in wafer A2513 at all electron densities because the surface is very far away from the 2DEG. For the 80 nm deep 2DEG in wafer A2512, even though surface charge causes less scattering than either background impurities and/or interface roughness, it is still strong enough to cause the mobility to be noticeably less than wafer A2513.}
    \label{FigAppndx-SCmobility}
\end{figure}

Charge can accumulate at the surface of semiconductors for a variety of reasons, be it from the local reorganization of the crystal lattice and bandstructure, redistribution of free charges (\textit{e.g.}, from ionized impurities), or the presence of excited states/dangling bonds to name a few. In GaAs/AlGaAs heterostructures at low temperatures, these surface charges are usually not mobile and their sheet density has been shown to be constant \cite{Kawaharazuka01}, consistent with the ``frozen surface model''. If the 2DEG is close to the surface (2DEG depth below the surface is $|d| \lesssim 100$ nm), these surface charges cause scattering to 2DEG carriers through Coulomb interactions \cite{Wendy10, WangDQ13}, as illustrated in Figure \ref{FigAppndx-SCmobility}.

Model-wise, surface charges are treated the same way as a delta-doped layer in a modulation-doped structure, located at the surface \cite{Ando82-A, Gold88}. Therefore the scattering potential $|U(q)|^2$ for surface charge becomes:
\begin{equation}
|U(q)|^2_{\textsc{sc}} = N_{\textsc{sc}}
\left(\frac{e^2}{2\epsilon_0\epsilon_r q}\right)^2
\frac{e^{-2q|d|}}{(1+q/b)^6}
\label{eq:U-SCsquare}
\end{equation}
\noindent where $N_{\textsc{sc}}$ is the surface charge sheet density and the rightmost fraction is the form factor $F_1(q,d)^2$ obtained from \cite{Ando82-A,Gold89-C, MacLeod09, Wendy10}:
\begin{eqnarray} 
F_i(q, z)&=&\int_0^\infty |\psi(z')|^2~e^{-q|z-z'|}~dz'\qquad\quad\qquad\label{eq:F(q,z)i}\\
F_1(q, d)&=&e^{-q|d|}\left(\frac{b}{b+q}\right)^3 = \frac{e^{-q|d|}}{(1+q/b)^3}
\label{eq:F1(q,z)}
\end{eqnarray}
\noindent where $z$ is the coordinate of the surface charge plane, and $z=d<0$ in equation (\ref{eq:F1(q,z)}). Substituting eqn.\,(\ref{eq:U-SCsquare}) into eqn.\,(\ref{eq:FermiGoldenRule}) and rewriting the integral in terms of $d\theta$ gives:
\begin{eqnarray}
\frac{1}{\tau_{\textsc{sc}}}=\frac{N_{\textsc{sc}}\,m^*}{2\pi\hbar^3 k_F^2}
\left(\frac{e^2}{2\epsilon_0\epsilon_r}\right)^2
\int_0^\pi \frac{e^{-2q|d|}}{\epsilon(q)^2\,(1+q/b)^6}~d\theta~.\qquad\label{eq:1byTauSC}
\end{eqnarray}

\begin{table}[t]
    \caption{List of single heterojunction wafers used in Fig.\,\ref{FigAppndx-SCmobility}, all grown with the same heterostructure as shown in Fig.\,\ref{Fig1}(a) [also see Ref.\,\onlinecite{Wendy10}], and their mobility fit parameters. Here, $\overline{N_{\textsc{bi}}} = N_{\textsc{bi-1}} = N_{\textsc{bi-2}}$ is assumed.}
    \begin{ruledtabular}
    \begin{tabular}{cccccc}
    Wafer & 2DEG & $\Delta$ & $\Lambda$ & $\overline{N_{\textsc{bi}}}$ & $N_{\textsc{sc}}$\\
    ID & depth (nm) & (nm) & (nm) & (cm$^{-3}$) & (cm$^{-2}$) \\
    \hline
    A2513& 310 & 0.19 & 18 & 1.3$\times$10$^{14}$ & $<$1$\times$10$^{10}$ \\
    A2512& \,~80 & 0.19 & 18 & 1.3$\times$10$^{14}$ & 1$\times$10$^{11}$ \\
    A2511& \,~40 & 0.19 & 18 & 1.3$\times$10$^{14}$ & 4$\times$10$^{11}$ \\
    error & $\pm$1 & $\pm$0.01 & $\pm$1 & $\pm$3\% & $\pm$3\%
    \end{tabular}
    \end{ruledtabular}
    \label{tab:Appndx-SCmobility}
\end{table}

\section{Background Impurities in AlGaAs}

Impurity atoms are invariably incorporated into semiconductor heterostructures during MBE growth. These can be either intentional dopants (for modulation doping) or non-intentional dopants (background impurities). In the dopant-free wafers considered here, only non-intentional background impurities are present, characterized by a volume impurity concentration $N_{\textsc{bi}}$. Ionized impurity scattering tends to dominate over other forms of scattering at very low carrier densities (\textit{e.g.}, interface roughness scattering or alloy scattering). Background impurity scattering from the AlGaAs and GaAs layers are treated separately in our model.

To quantify Coulomb scattering from impurities in the AlGaAs barrier, eqn.\,(\ref{eq:U-SCsquare}) for a delta-doped layer is integrated over the AlGaAs barrier volume (semi-infinite layer approximation), replacing $|d|$ with $|z|$, yielding the following scattering potential:
 \begin{eqnarray}
|U(q)|^2_{\textsc{bi-1}}&=&\left(\frac{e^2}{2\epsilon_0\epsilon_r q}\right)^2
\int_{-\infty}^0 N_{2D}^{\text{\tiny AlGaAs}}~ \frac{e^{-2q|z|}}{(1+q/b)^6}~dz~\qquad \\
&=&N_{\textsc{bi-1}}\left(\frac{e^2}{2\epsilon_0\epsilon_r q}\right)^2
F_{\text{\tiny AlGaAs}}(q)
\label{eq:U-AlGaAs}
\end{eqnarray}
where $N_{2D}^{\text{\tiny AlGaAs}}$ is the 2D sheet concentration of impurities in the AlGaAs layer, $N_{\textsc{bi-1}}$ is the volume impurity concentration in the AlGaAs layer, and $F_{\text{\tiny AlGaAs}}(q)$ is:
\begin{equation}
F_{\text{\tiny AlGaAs}}(q) = \int_0^\infty F_1(q,z')^2~dz' = \frac{1}{2q(1+q/b)^6}~.
\end{equation}
\noindent where $F_1(q,z')$ is defined in eqn.\,(\ref{eq:F1(q,z)}) with $z'>0$. Substituting eqn.\,(\ref{eq:U-AlGaAs}) into eqn.\,(\ref{eq:FermiGoldenRule}), simplifying and rewriting the integral in terms of $d\theta$ gives:
\begin{eqnarray}
\frac{1}{\tau_{\textsc{bi-1}}}&=&\frac{N_{\textsc{bi-1}}\,m^*}{2\pi \hbar^3 k^2_F}
\left(\frac{e^2}{2\epsilon_0\epsilon_r}\right)^2
\int^\pi_0 \frac{F_{\text{\tiny AlGaAs}}(q)}{\epsilon(q)^2}\,d\theta~.\qquad
\label{eq:1byTauBI-AlGaAs}
\end{eqnarray}

\section{Background Impurities in GaAs}
\label{Appdx:HEMT-fits}

Similarly to the treatment above for the AlGaAs layer, the $|U(q)|^2$ term for scattering from a strictly 2D charge layer (zero thickness) in the GaAs layer is:
\begin{equation}
|U(q)|^2_{2D} = N^{\text{\tiny GaAs}}_{2D}
\left(\frac{e^2}{2\epsilon_0\epsilon_r q}\right)^2~F_2(q,z)^2
\label{eq:U-2D-GaAs}
\end{equation}
\noindent where $N_{2D}^{\text{\tiny GaAs}}$ is the 2D sheet density of impurities in the GaAs layer, $z$ is the coordinate of the 2D charge plane, and eqn.\,(\ref{eq:F(q,z)i}) is used to calculate the form factor $F_2(q,z)$ for a 2DEG interacting with a 2D charge layer located in the same GaAs layer \cite{SternHoward67,Ando82-A}:
\begin{subequations}
\begin{eqnarray} 
F_2(q, z)|_{q\,=\,b}&=&\frac{1+2bz+2b^2z^2+\frac{4}{3}b^3z^3}{8\,e^{qz}} \\
F_2(q, z)|_{q\,\neq\, b}&=&\frac{e^{(b-q)z}-(c_0+c_1z+c_2z^2)}{(1-q/b)^3\,e^{bz}}
\qquad
\end{eqnarray}
\end{subequations}
\begin{eqnarray}
\text{where}\qquad &&c_0=\frac{2q(3b^2+q^2)}{(b+q)^3}\qquad\quad \nonumber \\
&&c_1=\frac{4bq(b-q)}{(b+q)^2} \nonumber \\
&&c_2=\frac{q(b-q)^2}{(b+q)} \nonumber 
\end{eqnarray}
\noindent where $z>0$. The expression for $F_2(q,z)$ is more complex than $F_1(q,z)$ owing to the direct overlap of the 2DEG wavefunction and the charge layer.

\begin{figure}[t]
    \includegraphics[width=\columnwidth]{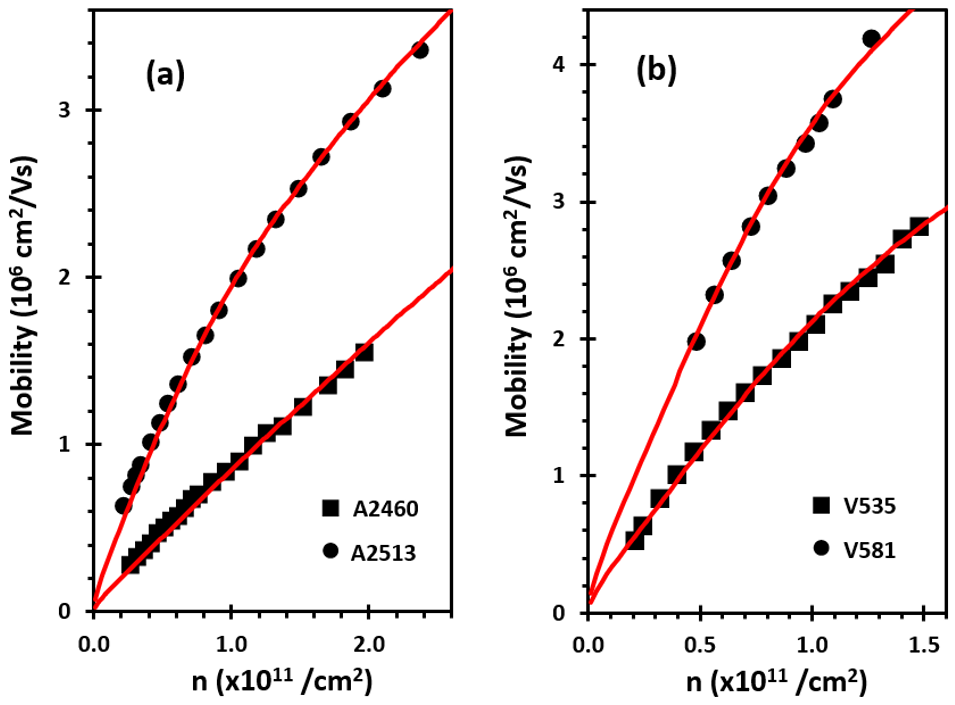}
    \caption{(color online) One-parameter mobility fit of 2DEGs, showcasing typical MBE chamber clean-up within a growth campaign, resulting in much-reduced background impurity concentrations. Electron mobilities, at $T$\,=\,1.5\,K in the dark, of 310 nm deep 2DEGs before (squares) and after (circles) the growth of some $\sim$\,50 wafers in the: (a) `A' chamber and (b) `V' chamber. Experimental data taken from Ref.\,\onlinecite{WendyPhD}. Solid (red) lines are fits from the model in section III with the parameters listed in Table \ref{tab:AppndxWafers}. The average background impurity concentration $\overline{N_{\textsc{bi}}}$ dropped from 3.3$\times$10$^{14}$ /cm$^3$ in wafer A2460 to 1.3$\times$10$^{14}$ /cm$^3$ in wafer A2513 in chamber `A', and from 1.3$\times$10$^{14}$ /cm$^3$ in wafer V535 to 0.7$\times$10$^{14}$ /cm$^3$ in wafer V581 in chamber `V'.}
    \label{FigAppndx-MBEcleanup}
\end{figure}  

To quantify Coulomb scattering from impurities in the GaAs layer, eqn.\,(\ref{eq:U-2D-GaAs}) is integrated over the GaAs layer (assuming a semi-infinite layer), yielding this scattering potential: 
\begin{eqnarray}
|U(q)|^2_{\textsc{bi-2}}&=&\left(\frac{e^2}{2\epsilon_0\epsilon_r q}\right)^2
\int_0^\infty N_{2D}^{\text{\tiny GaAs}}~F_2(q,z)^2~dz\qquad\qquad \\
&=&N_{\textsc{bi-2}}
\left(\frac{e^2}{2\epsilon_0\epsilon_r q}\right)^2 F_{\text{\tiny GaAs}}(q)
\label{eq:U-GaAs}
\end{eqnarray}
where $N_{\textsc{bi-2}}$ is the volume impurity concentration in the GaAs layer, and the form factor $F_{\text{\tiny GaAs}}(q)$ is defined as:
\begin{equation}
F_{\text{\tiny GaAs}}(q)=
\int_0^\infty F_2(q,z)^2~dz\qquad\qquad\qquad\qquad
\label{eq:FormFactor-GaAs}
\end{equation}
\begin{subequations}
\begin{eqnarray}
F_{\text{\tiny GaAs}}(q)|_{q\,=\,b} ~&=&~\frac{69}{128\,q}  \\
F_{\text{\tiny GaAs}}(q)|_{q\,\neq\, b}~&=&~\frac{1}{2q}\,\frac{1}{(1-q^2/b^2)^6}
\bigg(1+6\frac{q}{b}-33\frac{q^2}{b^2} \nonumber \\
&&\quad +\frac{75}{2}\frac{q^3}{b^3}+15\frac{q^4}{b^4}-36\frac{q^5}{b^5}-15\frac{q^6}{b^6}
\nonumber \\
&&\quad +33\frac{q^7}{b^7}-10\frac{q^9}{b^9}+\frac{3}{2}\frac{q^{11}}{b^{11}}\bigg).
\label{eq:F-GaAs}
\end{eqnarray}
\end{subequations}

Substituting eqn.\,(\ref{eq:U-GaAs}) into eqn.\,(\ref{eq:FermiGoldenRule}) and simplifying/rewriting the integral in terms of $d\theta$ gives:
\begin{eqnarray}
\frac{1}{\tau_{\textsc{bi-2}}}&=&\frac{N_{\textsc{bi-2}}\,m^*}{2\pi \hbar^3 k^2_F}
\left(\frac{e^2}{2\epsilon_0\epsilon_r}\right)^2
\int^\pi_0 \frac{F_{\text{\tiny GaAs}}(q)}{\epsilon(q)^2}\,d\theta~.\qquad \label{eq:1byTauBI-GaAs}
\end{eqnarray}

In the case of deep 2DEGs (whose depth below the surface is greater than 300 nm), curve-fitting the mobility can be reduced to a single free variable (assuming experimentally-determined interface roughness terms $\Delta$ and $\Lambda$): the average background impurity concentration by setting $\overline{N_{\textsc{bi}}} = N_{\textsc{bi-1}} = N_{\textsc{bi-2}}$. Single-parameter fits of four 2DEG mobilities as a function of electron density are shown in Figure \ref{FigAppndx-MBEcleanup}.

\begin{table}[t]
    \caption{List of single heterojunction wafers used in Figure \ref{FigAppndx-MBEcleanup}, grown in two different MBE chambers with the same heterostructure as shown in Fig.\,\ref{Fig1}(a), and the mobility fit parameters for each wafer. Here, $\overline{N_{\textsc{bi}}} = N_{\textsc{bi-1}} = N_{\textsc{bi-2}}$ is assumed.}
    \begin{ruledtabular}
    \begin{tabular}{ccccc}
    Wafer & 2DEG & $\Delta$ & $\Lambda$ & $\overline{N_{\textsc{bi}}}$ \\
    ID & depth (nm) & (nm) & (nm) & (cm$^{-3}$) \\
    \hline
    A2460& 310 & 0.20 & 20 & 3.3$\times$10$^{14}$ \\
    A2513& 310 & 0.19 & 18 & 1.3$\times$10$^{14}$ \\
    V535 & 310 & 0.15 & 16 & 1.3$\times$10$^{14}$ \\
    V581 & 310 & 0.14 & 16 & 0.7$\times$10$^{14}$ \\
    error & $\pm$1 & $\pm$0.01 & $\pm$1 & $\pm$3\%
    \end{tabular}
    \end{ruledtabular}
    \label{tab:AppndxWafers}
\end{table}
\begin{table}[t]
    \caption{List of model parameters for wafers W639, W640, and W641 for mobilities in the dark (before illumination), when allowing $N_{\textsc{bi-1}} (\text{AlGaAs}) \neq N_{\textsc{bi-2}}$ (GaAs). The fits suggest AlGaAs has more impurities than GaAs, $\frac{N_{\textsc{bi-1}}}{N_{\textsc{bi-2}}}= 2.75$.}
    \begin{ruledtabular}
    \begin{tabular}{ccccccc}
    Wafer & $d$ & $\Delta$ & $\Lambda$
    & $N_{\textsc{bi-1}}$ & $N_{\textsc{bi-2}}$ & $N_{\textsc{sc}}$\\
    ID & (nm) & (nm) & (nm) & (cm$^{-3}$) & (cm$^{-3}$) & (cm$^{-2}$) \\
    \hline
    W639 & 160 & 0.15 & 14
    & 3.3$\times$10$^{14}$ & 1.2$\times$10$^{14}$ & $<$1$\times$10$^{10}$ \\
    W640 & 110 & 0.11 & 14
    & 3.3$\times$10$^{14}$ & 1.2$\times$10$^{14}$ & 0.2$\times$10$^{11}$ \\
    W641 & \,~60 & 0.11 & 14
    & 3.3$\times$10$^{14}$ & 1.2$\times$10$^{14}$ & 1.7$\times$10$^{11}$ \\
    error & $\pm$1 & $\pm$0.01 & $\pm$1 & $\pm$3\% & $\pm$3\% & $\pm$3\%
    \end{tabular}
    \end{ruledtabular}
    \label{tab:Appndx-GaAs-AlGaAs-ratio}
\end{table}

\section{Relaxing $N_{\textsc{bi-1}} = N_{\textsc{bi-2}}$}
\label{Appdx:Alternate-fits}

Table \ref{tab:Appndx-GaAs-AlGaAs-ratio} lists fit parameters to Series I wafers when the constraint $N_{\textsc{bi-1}} = N_{\textsc{bi-2}}$ is relaxed. As also found in Ref.\,\onlinecite{Wendy10} and Ref.\,\onlinecite{MacLeod09}, the ratio $N_{\textsc{bi-1}}/N_{\textsc{bi-2}}$ providing the best fit to the data (all else being equal) was found to be approximately $\sim$\,3, meaning there are about 3$\times$ more charged impurities in AlGaAs than in GaAs. Al atoms are more reactive than Ga atoms, and are thought to getter more impurities (most likely oxygen) \cite{Manfra14}.

\end{document}